**TITLE**

An approach for the identification of targets specific to bone metastasis using cancer genes interactome and gene ontology analysis

**AUTHORS**

Shikha Vashisht and Ganesh Bagler*

**AFFILIATIONS**

Biotechnology Division, CSIR-IHBT, Institute of Himalayan Bioresource Technology, Council of Scientific and Industrial Research, Palampur, 176061 (H.P.), India.

*Author for correspondence**:** Phone Number: +919816931177; Fax: +91-1894-230433; Email: bagler@ihbt.res.in, ganesh.bagler@gmail.com.

**ABSTRACT**

Metastasis is one of the most enigmatic aspects of cancer pathogenesis and is a major cause of cancer-associated mortality. Secondary bone cancer (SBC) is a complex disease caused by metastasis of tumor cells from their primary site and is characterized by intricate interplay of molecular interactions. Identification of targets for multifactorial diseases such as SBC, the most frequent complication of breast and prostate cancers, is a challenge. Towards achieving our aim of identification of targets specific to SBC, we constructed a 'Cancer Genes Network', a representative protein interactome of cancer genes. Using graph theoretical methods, we obtained a set of key genes that are relevant for generic mechanisms of cancers and have a role in biological essentiality. We also compiled a curated dataset of 391 SBC genes from published literature which serves as a basis of ontological correlates of secondary bone cancer. Building on these results, we implement a strategy based on generic cancer genes, SBC genes and gene ontology enrichment method, to obtain a set of targets that are specific to bone metastasis. Through this study, we present an approach for probing one of the major complications in cancers, namely, metastasis. The results on genes that play generic roles in cancer phenotype, obtained by network analysis of 'Cancer Genes Network', have broader implications in understanding the role of molecular regulators in mechanisms of cancers. Specifically, our study provides a set of potential targets that are of ontological and regulatory relevance to secondary bone cancer.

**INTRODUCTION**

Cancer is a disease of multiple systems and components that interact at both molecular and cellular levels leading to initiation, progression and spread of the disease [1,2]. The changing interactions of these systems in a dynamic environment underscore the inherent complexity of the disease. Until recently, cancer has been studied with a reductionist approach focusing on a specific mutation or a pathway. Lately there has been a tremendous increase in systems-level study of cancer and the use of integrative approaches to understand mechanisms of cancers [3,4] and their metastases [5,6].

Metastasis is one of the most enigmatic hallmarks of cancers characterized by complex molecular interactions [1,7]. It is responsible for as much as 90% of cancer-associated mortality, yet remains the most poorly understood component of cancer pathogenesis [7,8]. Tumor metastasis is a multistage process during which malignant cells spread from the primary tumor to discontiguous organs [7]. Metastatic dissemination involves a sequence of steps involving invasion, intravasation, extravasation, survival, evasion of host defense and adaptation to the foreign microenvironment [7,8].

Secondary bone cancer (SBC) is a complex disease involving interplay of osteolytic and osetoblastic mechanisms [9] (Figure 1). Bone metastases are the most frequent complication of breast and prostate cancers with a very high propensity of metastasizing to bone causing bone pain, fracture, hypercalcemia and paralysis [10–13]. Breast and prostate carcinomas are often known to take years to develop metastatic colonies (in a limited number of sites) suggesting that in these cancers, cells employ distinct adaptive programs to laboriously cobble together complex shifts in gene-expression programs [14]. Many molecules and associated pathways are reported to be involved in metastasis of cancer cells from breast cancer [13,15–22] and those from prostate cancer [10,18,23–30].

Cancers are characterized by hallmark processes and shared mechanisms involved in expression of disease phenotype. It is a challenge to identify such genes involved in generic cancer mechanisms. Identification of such 'generic cancer genes' may help us focus on 'disease specific cancer genes' of potential therapeutic value. Due to complexity and subtle mechanisms involved in metastasis, it is difficult to identify their control mechanisms. Therefore it is important to have methods for identification of genes and regulatory mechanisms that are key to a complex pathogenic state such as secondary bone cancer. Complex network models of interactomes, along with graph-theoretical analysis and overrepresentation studies, present us a useful strategy for probing molecules that are central to SBC mechanisms and hence potential therapeutic targets.

Cellular functions reflect the state of the cell as a function of an intricate web of interactions among large number of genes, metabolites, proteins and RNA molecules. A disease phenotype reflects various pathobiological processes that interact in a complex network and is rarely a consequence of abnormality in a single effector gene product [3]. Understanding diseases in the context of organizing principles of the architecture of biological networks allows us to address some fundamental properties of genes that are involved in disease. Study of disease protein interactomes offer a better understanding of disease-specific genes and processes involved and may offer better targets for drug development. Molecular interaction networks are characterized by the presence of a few highly connected nodes, often called hubs, suggesting a special role of these promiscuous interactors. Hubs of protein interactomes are more likely to be essential for the survival [31] and also reported to be important for cellular growth rate [32]. Proteins with high betweenness [33,34] are reported to have much higher tendency to be essential genes [35,36]. Cancer proteins are reported to be more central in the protein interactome and are, on an average, involved in twice as many interactions as those of non-cancer proteins [37].

The Gene Ontology [38] project provides an ontology of defined terms representing gene product properties. The ontology covers three domains: cellular component, parts of a cell or its extracellular environment; molecular function, elemental activities of a gene product at the molecular level and biological process, sets of molecular events with a defined beginning and end, pertinent to the functioning of integrated living units. GO enrichment methods provide a way to extract biological insight from a set of genes using the power of gene sets [39]. Enrichment analysis involves identification of GO terms that are significantly overrepresented in a given set of genes using statistical models such as hypergeometric and chi squared distributions [40]. A large repertoire of tools has been developed in recent years for enrichment analysis [41,42]. Methods of network analysis and enrichment studies have been effectively used to identify targets of diseases such as chronic fatigue syndrome [43], major depressive disorder [44], glioblastoma [45], colorectal carcinogenesis [46] and primary immunodeficiency disease [47].

In this study, we aimed to identify secondary bone cancer specific genes. Towards this goal, we used a composite strategy (Figure 2) involving identification of cancer genes specifying generic cancer mechanisms, compilation of genes implicated in metastasis to bone and identification of genes annotated with GO terms specific to secondary bone cancer, to obtain disease specific targets. While network analysis provides a systems perspective of complex molecular mechanisms and helps to identify its central components (functional elements), gene enrichment method enables identification of characteristic ontological features of the gene sets. We first constructed a representative protein interactome of all cancer genes and obtained hubs that are involved in generic cancer mechanisms. Further, we compiled a set of experimentally verified genes from the literature that is involved in metastasis of primary breast and prostate cancer into bone, the dominant cause of secondary bone cancer. Using a combination of protein interactome analysis and gene ontology enrichment studies, we obtained a set of genes (targets) specific to SBC mechanisms. Our study provides an approach to identify targets specific to a complex disease phenotype (bone metastasis) by combining systems-level interactome analysis and ontological studies.

**RESULTS**

**CGN as a representative interactome of cancer mechanisms**

We intended to construct an interactome that represents mechanisms involved in processes contributing to cancers. For this purpose we used genes listed in CancerGenes database [48], a compilation of cancer genes that are causally implicated in oncogenesis. We obtained 3164 cancer genes from CancerGenes database, which were used to construct an interactome. These genes were mapped on Human Protein Reference Database [49], a database of curated proteomic information pertaining to human proteins, to construct the Cancer Genes Network (CGN). CGN thus represents an intricate network of cancer proteins. CGN comprises 11602 interactions among 2665 proteins. Figure 3 depicts the CGN, a representative protein interactome of molecular agents and their regulatory interactions, involved in disease phenotype of cancers. Based on earlier reports [3], we hypothesize that proteins that are key to the structural integrity and interaction dynamics of CGN would correspond to proteins involved in regulatory mechanisms generic to cancers.

**Topologically central genes of CGN correlate to generic cancer mechanisms**

Molecular interaction networks have been reported to have a scale-free nature marked by the presence of few hubs that are critical for the networks [50]. Such hubs of protein interactomes are reported to be more essential for the survival [31] and also important for cellular growth rate [32]. Proteins with high betweenness are reported to have much higher tendency to be essential genes [35,36]. Cancer proteins are reported to be more central in the protein interactome and are, on an average, involved in twice as many interactions as non-cancer proteins [37]. As reported for other biological molecular networks [50,51], we find that CGN has a scale-free nature indicating presence of exceptionally promiscuously interacting hub nodes and those mediating a large number of interactions (Figure 4) [52].

We computed seven parameters that reflect topological features (degree [53] and neighborhood connectivity [54]), network flow (betweenness [33,34], stress [33,34] and average shortest path length [55]) and local clustering (clustering coefficient [53,55,56] and topological coefficient [57]). The 'key' proteins, that are topologically and dynamically central to CGN, were identified by network analysis. For this purpose 'degree', 'stress' and the normalized counterpart of the latter, 'betweenness', were used for identification of central proteins of CGN. We find that the selected parameters show very high mutual positive correlations (Figure 5). The best-ranked proteins from each of these parameters were compiled, for various thresholds (Top25—Top200), to identify proteins designated as 'hubs' (Set-A in Figure 2 and Supplementary Information Table S1.1).

We find that the 'hubs' (Figure 3; Top75 threshold), thus identified, include genes involved in regulation of processes known to be generically present in most cancers [1]. Proteins known to be involved in growth signals, thus leading to self-sufficiency in cancer cells, such as TGFBR1, EGFR, IGFR1R, GRB2, are in the 'hubs' of CGN [2]. The hubs also include RB1, BCL2, AKT1, CDK2, CASP8 which are involved in mechanisms of apoptosis, to which cells are known to acquire resistance in all types of cancers [1,58]. The TP53 tumor suppressor protein, known to be involved in most commonly occurring loss of proapoptotic regulation and affecting the apoptotic effector cascade [59,60], is also present in the CGN hubs. One of the hub proteins, SMAD4, is known to be involved in differentiation, apoptosis and cell cycle [61]. MYC, known to upregulate cyclins and downregulate CDKN1A (P21), is also one of the hubs of CGN.

Further, we checked how well the identified hubs correlate with KEGG-PIC [62], a collection of genes from generic pathways involved in cancers (Supplementary Information Table S1.2). Figure 6 shows the overlap of hub genes (Top25 to Top200) with KEGG-PIC genes. We find that indeed the Top25 hubs, comprising the most promiscuously interacting proteins and mediating a large number of interactions in cancer genes interactome, have a 74% overlap with KEGG-PIC. The precision of identification of generic cancer proteins drops as the definition of 'hubs' is relaxed further (57% for Top50 and 52% for Top75). As a negative control, for the ability of network-metrics-based ranking to identify generic proteins, we use genes ranked worst according to the selected parameters. We find that these genes have a very poor overlap with KEGG-PIC genes, even when compared to that expected from corresponding random samplings (Figure 6). We treat Top25, Top50 and Top75 CGN hubs as representatives of generic cancer genes and use them, further, to identify SBC-specific targets.

**GO enrichment of CGN hub genes**

We performed GO enrichment analysis of CGN 'hub' genes identified. We expect the 'hubs' to represent generic cancer processes which are shared by most, and perhaps, all types of human cancers. The molecular machinery regulating proliferation, differentiation and death of all mammalian cells is highly similar [1]. The genetic transformation of normal body cells results in defects of regulatory circuits that govern normal cell proliferation and homeostasis, and collectively dictate malignant growth.

Cancer cells show self-sufficiency of growth signals involving alteration of transcellular or intracellular mechanisms [2]. The growth signaling pathways are suspected to suffer deregulation in all human tumors [2]. This was reflected in the following significantly overrepresented GO terms of CGN hubs: cellular response to growth hormone stimulus (GO:0071378), regulation of epidermal growth factor receptor signaling pathway (GO:0042058), transforming growth factor beta receptor signaling pathway (GO:0007179), signal complex assembly (GO:0007172).

In contrast to normal cells, which respond to antigrowth signals, cancerous cells are insensitive to antigrowth signals due to disruption of pathways related to cell cycle clock. We find that our list of significantly enriched GO terms for CGN hubs had many terms associated with cell cycle processes: mitotic cell cycle G1/S transition checkpoint (GO:0031575), regulation of G1/S transition of mitotic cell cycle (GO:2000045), DNA damage response, signal transduction by p53 class mediator (GO:0030330), regulation of stress-activated MAPK cascade (GO:0007173).

In human carcinogenesis, all types of cancers are known to acquire resistance to apoptosis [58]. The overrepresented GO terms of CGN hubs underline this observation: release of cytochrome c from mitochondria (GO:0001836), activation of pro-apoptotic gene products (GO:0008633), cell-type specific apoptotic process (GO:0097285), cellular component disassembly involved in apoptotic process (GO:0006921), positive regulation of anti-apoptosis (GO:0045768).

In addition, we find following significantly enriched 'molecular function' terms: transforming growth factor beta receptor, pathway-specific cytoplasmic mediator activity (GO:0030618), receptor signaling protein tyrosine kinase activity (GO:0004716), ephrin receptor binding (GO:0046875), MAP kinase activity (GO:0004707), beta-catenin binding (GO:0008013), cytokine receptor binding (GO:0005126), androgen receptor binding (GO:0050681).

All the GO-terms mentioned above broadly support generic cancer processes. The p-values for these 'significantly enriched GO terms of CGN Top75 hubs' are in the range of 10e-4 and 10e-13. Figure 7 depicts these overrepresented GO terms and statistics of their prevalence.

**Topologically central genes of CGN correlate with biological essentiality**

We further checked the biological relevance of the central genes of CGN. We divided CGN into two sets, essential and non-essential, using Mammalian Phenotype Ontologies [63] to classify genes as essential when they caused embryonic, perinatal, postnatal or neonatal lethality in mouse models [64] using phenotypic data from Mouse Genome Informatics (MGI) [65]. The percentage of essential genes were computed in the 'hubs' and their corresponding non-hubs for varying hub definitions. Figure 8 shows the statistics for percentage of hub and non-hub genes annotated as essential. Among the seven parameters computed, the hubs identified using degree, betweenness and stress have a significant percentage of essential genes (between 82% to 92%) (Figure 8A). Incidentally, none of the remaining four network parameters show good correlation with biological essentiality. The non-hubs, as expected from the random samplings statistics, have around 50% of essential genes associated, regardless of the metric used for ranking (Figure 8B). For hubs comprising of Top75 genes, the chosen parameters have a very high proportion of essential genes (between 88% and 89%), compared to the rest of the parameters. Figure 3 depicts the central nature of Top75 hub genes of CGN and highlights the essential and non-essential genes among them. This emphasizes the relevance of the network metrics chosen in encoding biological relevance. This is consistent with earlier reports about hubs being more essential [64] and network centrality of cancer genes [37].

**SBCGs and their characteristic GO terms**

Knowing that metastasis into bone is primarily caused by spread from primary prostate and breast cancers [10–13], we aimed at compiling genes linked with these processes. We compiled a list of 391 Secondary Bone Cancer (SBC) susceptible genes from literature reporting experimental studies (Set-B in Figure 2 and Supplementary Information Table S1.3). For identifying the relevance of these genes in the context of metastasis into bone, we performed an overrepresentation analysis using CGN as universe. Gene ontology enrichment studies of SBCGs are expected to reflect GO categories of biological processes and molecular functions that are relevant for metastasis into bone, consistent with the known aspects of regulatory mechanisms of bone metastasis (Figure 1).

We identified 93 significantly enriched GO terms of SBCGs for biological processes, molecular functions and cellular components. We identify CGN genes annotated with either of these 93 GO terms to construct 'Enriched Genes' set (Set-C in Figure 2). We find that out of the 93 significantly enriched GO terms, 31 are associated with metastasis or bone processes. Out of these 31 GO terms, 21 are most relevant for metastasis mechanisms and 10 for bone-related processes (Figure 9).

The following GO terms, that were significantly enriched, are related to bone processes: osteoblast differentiation (GO:0001649), regulation of bone remodeling (GO:0046850), endochondral ossification (GO:0001958), replacement ossification (GO:0036075), bone mineralization (GO:0030282), ossification (GO:0001503), cartilage development (GO:0051216), response to vitamin D (GO:0033280), collagen metabolic process (GO:0032963), collagen fibril organization (GO:0030199).

The following biological processes GO terms, that were significantly enriched, are related to mechanisms of metastasis: leukocyte migration (GO:0050900), cell-cell adhesion (GO:0016337), angiogenesis (GO:0001525), positive regulation of cell adhesion (GO:0045785), negative regulation of cell adhesion (GO:0007162), positive regulation of leukocyte migration (GO:0002687), regulation of leukocyte chemotaxis (GO:0002688), positive regulation of leukocyte chemotaxis (GO:0002690), positive regulation of catenin import into nucleus (GO:0035413).

Following molecular functions GO terms that were significantly enriched, are related to mechanisms of metastasis: laminin binding (GO:0043236), fibronectin binding (GO:0001968), fibroblast growth factor receptor binding (GO:0005104), platelet-derived growth factor binding (GO:0048407), extracellular matrix binding (GO:0050840), collagen binding (GO:0005518), integrin binding (GO:0005178), glycosaminoglycan binding (GO:0005539), cytokine activity (GO:0005125), carbohydrate binding (GO:0030246).

**Predicted SBC-specific targets**

For predicting SBC-specific targets of potential therapeutic value, we logically juxtaposed the three sets of genes obtained: CGN hub genes (Top75), SBCGs and the cancer genes annotated with GO terms overrepresented for SBCGs (Figure 2 and Figure 10). On the basis of specificity for secondary bone cancer, as depicted in Figure 10A, the Venn diagram corresponding to these gene sets was divided into seven distinct regions: Set-a, Set-b, Set-c, Set-ab, Set-bc, Set-ac and Set-abc. Overall, the sets belonging to central genes of CGN (Set-a, Set-ab, Set-ac and Set-abc) contain genes that are involved in generic cancer mechanisms (Figure 6) and are also essential (Figure 8). Hence, while obtaining SBC-specific targets, we don't consider these gene sets. Out of the rest of three sets, Set-b comprises genes that are already reported to be associated with SBC mechanisms, hence may not reveal novel genes. We focus on Set-c and Set-bc, corresponding to non-generic cancer genes annotated with SBC-specific GO terms and those reported to be having a role in SBC mechanisms, respectively, for the search of novel SBC-specific targets.

Set-bc and Set-c contain 53 and 134 genes, respectively. These lists were refined to select only those genes that are annotated with GO terms most relevant for SBC (Figure 9 and Supplementary Information Table S1.4), to obtain 28 and 60 genes from Set-bc and Set-c, respectively. We find that, in these sets, we were still left with KEGG-PIC genes known to be involved in generic cancer pathways. The list of targets was refined further to obtain a total of 72 targets that are non-generic to cancers and relevant for secondary bone cancer.

Out of these 72 genes, 14 genes are annotated with GO terms relevant to bone processes, 51 with metastasis and 7 with both (Figure 3). We predict these seven proteins, with ontological relevance to bone processes as well as metastasis, to be most potent targets and key regulators of metastasis into bone. The non-generic SBC-specific targets, with relevance to metastasis and bone mechanisms, identified are: SPP1 (Secreted Phospoprotein 1), CD44 (Cluster of Differentiation 44), CTGF (Connective Tissue Growth Factor), TNXB (Tenascin X), BMP1 (Bone Morphogenetic Protein 1), BMPR1A (Bone Morphogenetic Protein Receptor, Type IA) and VWF (Von Willebrand Factor).

We find that the SBC-specific targets identified are indeed relevant for the metastasis into bone and involved in regulation of SBC mechanisms (Supplementary Information Table S1.5). SPP1 is reported to interact with CD44 receptor and is thought to exert pro-metastatic effects leading to tumor progression by regulating the cell signaling events [66,67]. CD44 is also reported to enhance integrin-mediated adhesion and transendothelial migration of breast cancer cells [68]. CTGF expression has been shown to be associated with tumor development and progession. The overexpression of CTGF in breast cancer cells may promote their metastasis to bone [69]. BMPR1A, belonging to the family of transmembrane serine/threonine kinases, is reported to be necessary for extracellular matrix deposition by osteoblasts, while not essential for osteoblast formation or proliferation [70]. BMP1, which does not belong to TGFβ superfamily unlike other BMPs, is known to induce bone and cartilage development. TNXB is an extracellular matrix glycoprotein expressed in connective tissues including skin, joints and muscles and in known to have role in cell-matrix and cell-cell adhesion. It has been reported that TNX deficiency leads to the invasion and metastasis of tumor cells by facilitating increase in the activity of MMPs which results in the degradation of laminin. The over-expression of TNX could be of potential therapeutic benefit in reducing tumor progression [71]. VWF is a large multimeric glycoprotein present in blood plasma and is produced constitutively in endothelium, megakaryocytes and subendothelial connective tissue. It has been reported that in osteosarcoma tumors the expression of VWF gets deregulated, potentially leading to metastasis [72].

**DISCUSSION**

Towards our goal of identifying key targets specific for mechanisms of metastases of primary breast and prostate cancers to bone, we used network analysis and gene ontology enrichment studies. The motivation behind using a combination of network analysis and gene enrichment methods is that, while the prior provides a systems perspective of complex molecular mechanisms and helps to identify its central components (functional elements), the latter enables identification of characteristic ontological features of the gene sets. Secondary Bone Cancer (SBC) is a complex disease triggered from the primary form of cancers, most commonly through breast cancer and prostate cancer. Many pathways and molecular regulators involved in the metastasis of primary prostate cancer and breast cancer are well known. Starting from SBC-related proteins and interactome of proteins involved in cancer mechanisms, using network analysis and overrepresentation studies, we identify targets that are specific to SBC.

The final list of seven targets was identified using generic cancer genes obtained with Top75 hubs. We repeated this task with Top25 and Top50 lists, which have better overlap with KEGG-PIC genes (74% and 57%, respectively), as compared to that of Top75 hubs (52%) (Figure 7). Using the procedure described, we identified SBC-specific targets starting from Top25 and Top50 CGN hubs. Figure 11 illustrates the results of these experiments obtained with better overlap with generic cancer genes (KEGG-PIC). Using these thresholds we obtained 31 and 60 hubs, respectively, which were filtered from the potential target set. This resulted in increase in number of potential targets in Set-c to 141 and 139, respectively, compared to those obtained with Top75 hubs (134). There was no change in the number of genes (53) obtained in Set-bc. Even with these criteria that correspond better with generic genes, the number of targets (that are annotated with GO terms relevant to metastasis and that of bone processes) does not increase, indicating the soundness of criteria used for identification of SBC-specific targets.

Network analysis has been shown to be a very useful and potent tool in understanding the disease phenotype and probing for therapeutic targets [3]. The identification of network parameters relevant to the question(s) being asked is an important task. We chose degree, betweenness and stress that are known to be important in the topology of the disease interactome and dynamic interplay of proteins involved [3]. These parameters were also found to have very good correlation with 'essential genes' (Figure 8). The significant terms emerging from overrepresentation analysis of CGN too support our choice of parameters (Figure 7).

Gene Ontology (GO) defines a set of functional terms related by parenthood relationships forming a directed acyclic graph. It produces sets of explicitly defined, structured vocabularies that describe biological processes, molecular functions and cellular components of gene products [38]. GO classification is expected to become an increasingly powerful tool for data analysis and functional predictions as the ontologies and annotations continue to evolve [40]. It is characterized by high quality manual curation, consistent annotation standards across species, and has the advantage of lesser bias as compared to domain specific classification schemes due to its comprehensive nature [40]. GO enrichment methods, which provide means of identification of significantly overrepresented GO terms, could be effectively used to get biological insights from a given set of genes [40]. One of the critical points in GO enrichment analysis is the selection of background set for getting the correct results. In this study, we use a pool of cancer genes (CGN) as universe, which serves the purpose of a meaningful background set of cancer mechanisms.

Cancer is a complex genetic disease characterized by intricate regulations among a diverse set of cancer genes. Hence, it is useful to have a detailed map of interactions that could help in probing hallmark regulatory modules and perhaps cancer-specific motifs. Here, we used CancerGenes Database (CGDb) [48] as a premise of cancer genes to construct a representative cancer interactome. This could potentially be improved by compiling a more exhaustive list of cancer genes. Though, we believe that the ultimate set of generic cancer genes, central to the topology of the cancer genes network, would not alter much, it may enrich the data and enable a more meaningful analysis of subtle regulatory features of cancer phenotype. It would be interesting to see whether and how the modules of CGN reflect the hallmarks of cancers. Also, while a few of the network centrality metrics for cancer genes may enumerate biological essentiality, it would be interesting to explore a larger parameter space to search for topological correlates of essentiality. We use KEGG-PIC genes as a prototype of generic cancer mechanisms. There is scope for improvement of this data by inclusion of generic genes through curation from landmark studies [1,2,73,74]. We believe that better metrics that embody structure and information dynamics over the cancer genes interactome could be developed that are more successful in elucidating the regulatory features characterizing molecular circuitries of cancers.

In view of the complex and subtly intertwined regulatory mechanisms of cancers, we modeled it as a network of protein interactions and aimed to identify generic cancer genes that specify hallmark features of cancers. We find that, indeed hubs of the cancer genes interactome that are central to the structural integrity and dynamical interplay of proteins, correlate with genes from generic pathways of cancers. We believe that the methodology presented here could be useful in obtaining cancer-specific targets of potential therapeutic value.

**MATERIALS AND METHODS**

**Cancer Genes Network (CGN)**

We used HPRD (Human Protein Reference Database) [49], one of the most comprehensive resources of human protein-protein interactions (PPIs), to construct a reference human protein interactome. HPRD is a manually curated human protein-protein interaction resource containing 36617 unique human PPIs and 9427 associated proteins (Release 9: April 13, 2010). It is one of the best resources of human PPIs, containing the largest number of binary non-redundant human PPIs, largest number of genes annotated with at least one interactor, and largest citations of PPIs curated [75].

The data of cancer genes involved in carcinogenesis were compiled from CancerGenes database [48] (as on May 2012). CancerGenes database is a compilation of cancer gene lists annotated by experts with information from key publicly available databases. Cancer associated genes are collected from various sources such as Cancer Map Pathways, Sanger Cancer Gene Census, Sanger Catalogue of Somatic Mutations in Cancer, reviews on cancer, Entrez queries and prostate cancer list. This data of 3164 cancer genes was used to compile a protein interactome representing molecular mechanisms of cancers. The interactions associated with proteins corresponding to these genes were collected from HPRD [49]. This network was called CGN (Cancer Genes Network), in which nodes represent cancer genes and edges represent experimentally validated interactions between a pair of genes *i* and *j*. Thus CGN is an intricate network of cancer proteins, comprising of 11602 interactions among 2665 proteins. It contains a giant cluster of 2376 proteins interlinked via 11590 interactions and the rest fragmented into minor clusters and isolated nodes (Figure 3). To adjudge the scale-free nature of degree and betweenness distributions of CGN, we tested the power-law hypothesis and estimated the parameters for these distributions with the technique based on maximum likelihood methods and the Kolmogorov-Smirnov statistic (Figure 4) [52].

**Secondary Bone Cancer Genes (SBCGs)**

We curated and compiled 391 genes involved in 'metastasis to bone from primary prostate and breast cancer' (167 and 230 genes, respectively), through literature mining (Set-B in Figure 2). These genes were called SBCGs (Secondary Bone Cancer Genes). We used following keywords to search for SBCGs: "secondary bone cancer genes from primary prostate cancer", "secondary bone cancer genes from primary breast cancer" (Pubmed) and "genes involved in metastasis of prostate cancer to bone", "genes involved in metastasis of breast cancer to bone" (Google Scholar). Supplementary Information Table S1.3 provides the details of SBCGs compiled from the literature.

**Gene Ontology analysis**

Gene Ontology (GO) enrichment or overrepresentation analysis allows one to identify characteristic biological attributes in a given gene set. It is based on the hypothesis that functionally related genes should accumulate in the corresponding GO category. We used GOrilla [41], a tool to identify enriched GO terms, to obtain biological attributes characterizing CGN hub genes as well as SBCGs. It uses the hypergeometric distribution to identify enriched GO terms in a given set of genes. We performed GO enrichment using 'two unranked lists of genes' mode, with an 'unranked target set' in the background of an 'unranked source set'. We identify 'significantly enriched GO terms', for Biological Process, Molecular Function and Cellular Components, with p-value cut-off of 0.001 and those having at most 100 genes associated with, in the source data (B≤100). The latter criterion is used to weed out terms those are too generic. The CGN Top75 hub genes (target) were enriched against all genes present in CGN (source). Similarly, the SBCGs (target) were enriched against CGN (source). These GO enrichment experiments help us obtain biological attributes that characterize the CGN hubs and those characterizing SBCGs, respectively, in the background of 'cancer genes' universe.

We performed the GO enrichment studies at three different p-values (0.01, 0.001 and 0.0001) for, both, SBCGs and CGN Top75 gene-set. We found that the significantly enriched GO terms obtained for BP, MF and CC were similar for p-values 0.01 and 0.001. For stricter p-value of 0.0001, we found that many of the GO terms relevant for secondary bone cancer and with generic role in cancer mechanisms, respectively, were lost. In the case of SBCGs, 14 BP and 4 MF disease-specific GO terms were lost at p-value 0.0001, including following key GO terms: angiogenesis (GO:0001525), regulation of bone remodelling (GO:0046850), bone mineralization (GO:0030282), collagen fibril organisation (GO:0030199), integrin binding (GO:0005178), laminin binding (GO:0043236), platelet derived growth factor binding (GO:0048407) and regulation of chemotaxis (GO:0050920). In the case of CGN Top75 gene-set analysis, 20 BP and 2 MF GO terms were lost, including following GO terms important for generic cancer mechanisms, at p-value 0.0001: cellular component disassembly involved in apoptotic process (GO:0006921), positive regulation of MAPK cascade (GO:0043410), positive regulation of cell cycle process (GO:0090068), regulation of G1/S transition of mitotic cell cycle (GO:2000045), mitotic cell cycle G1/S transition checkpoint (GO:0031575), growth hormone receptor signaling pathway (GO:0060396), response to fibroblast growth factor stimulus (GO:0071774) and transforming growth factor beta receptor signaling pathway (GO:0007179).

The value of 'B' (number of genes from the source set associated with a given GO term) was chosen to increase the 'signal' (specific GO terms) and to reduce 'noise' (non-specific GO terms). High values of 'B' increase noise by populating the enrichment results with non-specific GO terms; whereas small values of B reduce the signal by rejecting specific and relevant GO terms. We performed multiple experiments at different thresholds of B-values. (i) For B≤50, we missed 6 BP and 3 MF GO terms specific to SBC including the following: positive regulation of cell adhesion, angiogenesis (GO:0001525), cell-cell adhesion (GO:0016337), cytokine activity (GO:0005125), glycosaminoglycan binding (GO:0005539) and carbohydrate binding (GO:0030246). Similarly, in the case of ACGN Top75 gene-set, 12 BP and 3 MF relevant GO terms were lost including DNA damage response, signal transduction by p53 class mediator (GO:0030330), regulation of DNA replication (GO:0006275), positive regulation of cell cycle process (GO:0090068), positive regulation of MAPK cascade (GO:0043410), Ras protein signal transduction (GO:0007265) and response to UV (GO:0009411). (ii) For 50<B≤100, we got most of the relevant terms for both SBCGs and ACGN Top75 gene-set. (iii) For B>100, almost all the terms obtained were non-specific (noise). Hence, we chose B≤100 to maximize the relevant GO terms and to reduce the non-specific GO terms in the enrichment results.

**Protein interactome analysis**

We performed network analysis of CGN to compute various graph-theoretical metrics, using NetworkAnalyzer plugin of Cytoscape [76]. We computed seven network centrality parameters based on network connectivity (degree [53] and neighborhood connectivity [54]), network flow (betweenness [33,34], stress [33,34], average shortest path length [55]) and local clustering (clustering coefficient [53,55,56] and topological coefficient [57]).

Degree [53] corresponds to the number of nodes adjacent to a given node *v*, where adjacent means directly connected. The degree distribution *P*(*k*) of a network is then defined to be the fraction of nodes in the network with degree *k*. Thus if there are *N* nodes in total in a network and $n_k$ of them have degree *k*, we have *P*(*k*) = $n_k$/*N*.

'Neighbourhood connectivity' [54] of a node $v$ is defined as the average connectivity of all neighbors of $v$. For a node $v$ with $k_v$ number of neighbors, the neighborhood connectivity is defined as:

$$nc(v) \equiv \frac{\sum_{w \in W} k_w}{k_v}$$

Where *W* is the set of neighbors of $v$, and $k_w$ is the degree of each of the neighboring node.

'Stress' [33,34] and (its normalized counterpart) 'betweenness' [33,34] enumerate number of shortest paths from all pairs of vertices passing through the node of interest. In a graph *G = (V, E)* with *N* nodes, stress ($str(v)$) is defined as the total number of shortest paths passing through a node $v$:

$$str(v) \equiv \sum_{s \neq v \in N} \sum_{t \neq v \in N} \sigma_{st(v)}$$

Where, $\sigma_{st(v)}$ is the number of shortest paths from *s* to *t* that pass through vertex $v$. Betweenness is normalized (with the total number of shortest paths in graph *G*, $\sum_{\substack{s \neq v \neq t \in V \\ s \neq t}} \sigma_{st}$) value of stress. The higher the value of stress/betweenness, the higher is the relevance of the protein as a critical mediator of regulatory molecules and/or functional modules.

The average shortest path length [55] of a vertex $v$ in graph $G$, corresponds to the average of all the shortest paths between $v$ and the rest of the vertices. The average shortest path length of vertex $v$ is defined as:

$$asp(v) \equiv \sum_{w \neq v \epsilon V} \frac{dist(v,w)}{(N-1)}$$

Where *V* is the set of nodes in *G*, *dist(v,w)* is the shortest path from $v$ to $w$, and *N* is the number of nodes in *G*.

Clustering Coefficient $C_v$ [53,55,56] of a node $v$ is defined as:

$$C_v \equiv \frac{2e_v}{k_v(k_v - 1)}$$

Where $k_v$ is the number of neighbors of $v$ and $e_v$ is the number of connected pairs of nodes between all neighbors of $v$.

Topological Coefficient $tc(v)$ [57] of a node $v$ is defined as,

$$tc(v) \equiv \frac{\sum_{w \neq v \epsilon V} J(v,w)}{k_v}$$

Where *V* is the set of nodes in *G* and *J(v,w)* is the number of neighbors shared between the nodes $v$ to $w$, plus one if there is a direct link between $v$ and $w$.

We used degree, betweenness and stress metrics for identification of hubs of CGN. These metrics have been reported to be useful in identification of hubs of biological relevance [31,35,36] and those relevant to cancer [37].

**Identification of hub nodes and their controls**

First, we ranked the genes of CGN for each of the chosen parameters (degree, betweenness and stress). Then, we compiled eight 'hub gene-sets' (called Top25, Top50, so on till Top200), containing genes with ranks above the cut-off threshold, for each of the three parameters (Supplementary Information Table S1.1). Each of these hub gene-sets contains hub genes identified by either of the three parameters. Thus defined, the size of a hub gene-set may be up to three times the hub cut-off threshold, depending upon the similarity between the hubs identified by each of the parameters. Top25, Top50 and Top75 hub gene-sets of CGN contain 31, 60 and 92 hub genes, respectively. Figure 3 depicts the CGN hub genes identified using Top75 hubs criterion. As a negative control for these hub gene-sets, we identified the corresponding genes (from the bottom of ranked lists), with lowest ranking for the chosen parameters. As random controls, we randomly sample a corresponding number of genes from CGN (1000 instances each).

**Compilation of genes involved in generic cancer mechanisms (KEGG-PIC)**

Cancer cell mechanisms could be envisaged as an elaborate integrated circuit of intracellular signaling networks [1,2]. This map of molecular mechanisms could be represented as a combination of circuits and subcircuits with considerable cross talk among them [1]. We compiled a set of (328) representative genes known to be implicated in generic cancer mechanisms from cancer pathways/circuits (Supplementary Information Table S1.2). 'Pathways in Cancer' (PIC) (hsa05200) from KEGG PATHWAY database, are representative of generic cancer circuits. Here, we call this 'generic cancer genes' set KEGG-PIC [62]. KEGG-PIC comprise the following KEGG pathways: colorectal cancer (hsa05210), pancreatic cancer (hsa05212), thyroid cancer (hsa05216), acute myeloid leukemia (hsa05221), chronic myeloid leukemia (hsa05220), basal cell carcinoma (hsa05217), melanoma (hsa05218), renal cell carcinoma (hsa05211), bladder cancer (hsa05219), prostate cancer (hsa05215), endometrial cancer (hsa05213), small cell lung cancer (hsa05222), non-small cell lung cancer (hsa05223) and glioma (hsa05214).

**Mouse phenotype data**

For inferring the biological significance of the network parameters, we divided CGN into two sets, essential and non-essential, using the phenotypic information of the corresponding mouse ortholog [47,77]. It is assumed that human orthologs of mouse genes could be mapped onto each other for their function and biological essentiality. We considered the classes of embryonic, perinatal, neonatal or postnatal lethality in mouse models as lethal phenotypes, and the rest of the phenotypes as non-lethal ones. The human orthologs of murine genes were considered as essential, when the murine gene was annotated with one of the following phenotypes [63]: neonatal lethality (MP:0002058), embryonic lethality (MP:0002080), perinatal lethality (MP:0002081), postnatal lethality (MP:0002082), lethality-postnatal (MP:0005373), lethality-embryonic/perinatal (MP:0005374), embryonic lethality before implantation (MP:0006204), embryonic lethality before somite formation (MP:0006205) or embryonic lethality before turning of embryo (MP:0006206). The human-mouse orthology and mouse phenotype data was obtained from Mouse Genome Informatics [65] (May 2012). Out of a total 2665 cancer genes of CGN, 1315 (49.34%) were essential genes and the rest 1350 (50.66%) were non-essential. For each hub definition, from the hub-genes identified using each of the seven network metric, we compute the percentage of hubs that are essential (Figure 8A). We also found the percentage of essential genes in the corresponding non-hubs (Figure 8B).

**Venn diagram**

We logically juxtaposed the hubs of cancer genes network, curated bone cancer metastasis genes and cancer genes annotated with characteristic SBC GO terms to identify SBC-specific targets (Figure 2). Figure 10A illustrates this process highlighting the hubs of CGN (shaded area) that correspond to generic cancer genes and subsets potentially containing genes specific to SBC (hatched area). Figure 10B depicts the data when Top75 hubs of CGN are taken into consideration. For this data, there are 92 hubs of which 21 (Set-ab and Set-abc) happen to be common with SBCGs and 9 of those (Set-ac and Set-abc) are common to 'enriched genes'. Among the 'secondary bone cancer enriched cancer genes' 55 are common to SBGCs (Set-bc and Set-abc). We find that, for this data, there are 2 genes (Set-abc) that happen to be CGN hubs that are common to SBCGs and are also annotated with characteristic SBC GO terms. The logical juxtaposition results for the data of Top25 and Top50 hubs of CGN are depicted in Figure 11A and Figure 11B, respectively.

**Prediction of SBC-specific candidate genes**

We propose that the genes that are specific to secondary bone cancer mechanisms would be annotated with characteristic GO terms that are obtained from overrepresentation analysis of a literature curated list of genes implicated in metastasis to bone. From the 'secondary bone cancer enriched cancer genes', that serve as a 'source set of targets' we filtered the CGN hubs (Set-ac and Set-abc; shaded area) as they are generic to cancers (Figure 6) and found to be correlating with essential genes (Figure 8). Towards our aim of identifying SBC-specific targets, we focused on genes in Set-c and Set-bc (hatched area in Figure 10). We identified SBC-specific targets by refining these sets of genes to obtain genes that are annotated with any of the GO terms representing 'bone processes' as well as that of 'metastasis' (Figure 9) (Supplementary Information Table S1.4).

**SUPPORTING INFORMATION**
**Supplementary Material S1.**

In the supplementary information we present the details of hubs of CGN; the KEGG-PIC genes that serve as a reference set of generic cancer genes; secondary bone cancer genes that were curated and compiled; characteristic GO terms that were used to obtain SBC-specific targets and relevance of SBC-specific targets identified from experimentally validated studies. The supplementary information contains 5 tables, out of which Table S1.3 has 46 references, Table S1.4 has 23 references and Table S1.5 has 8 references.

**ACKNOWLEDGEMENTS**

We acknowledge the computational infrastructure provided by Institute of Himalayan Bioresource Technology (CSIR-IHBT), a constituent national laboratory of Council of Scientific and Industrial Research, India. The authors thank Dr. Paramvir Singh Ahuja for the encouragement and support. Authors thank Vinay Randhawa for technical help in manuscript preparation. The CSIR-IHBT communication number for this article is 2245.

**FIGURES:**

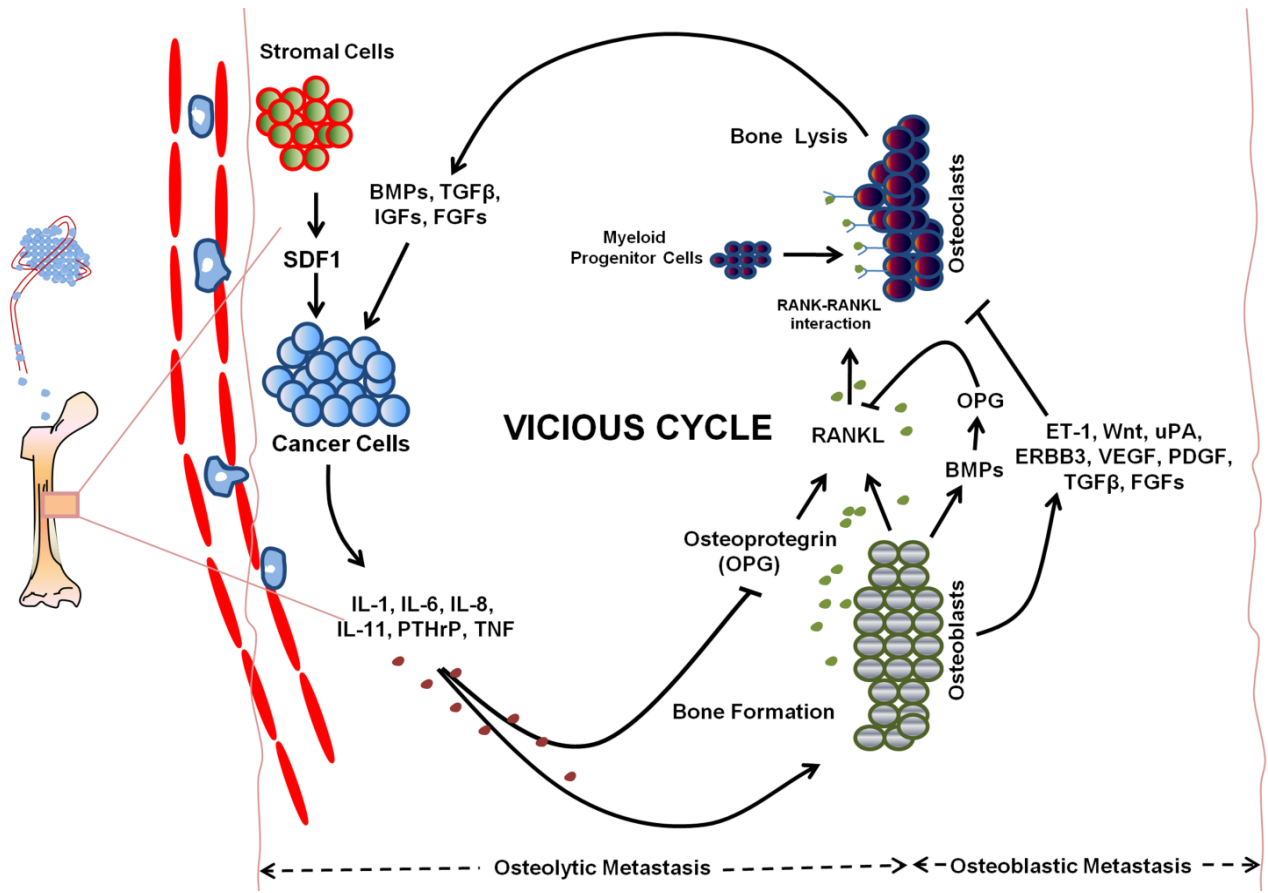

**Figure 1. Regulatory mechanisms underlying metastasis to bone reflecting complex interplay of molecules.**
Bone metastasis results from imbalance of normal bone remodeling process involving osteolytic (leading to bone destruction) and osteoblastic (leading to aberrant bone formation) mechanisms. Breast cancer metastases are usually osteolytic, whereas prostate cancer metastases are usually osteoblastic. **Osteolytic metastasis**: Osteolytic metastasis of tumor cells involves a "vicious cycle" between tumor cells and the skeleton. The vicious cycle is propagated by four contributors: tumor cells, bone-forming osteoblasts, bone resorbing osteoclasts and stored factors within bone matrix. Osteoclast formation and activity are regulated by the osteoblast, adding complexity to the vicious cycle. Tumor cells release certain factors including IL-1, IL-6, IL-8, IL-11, PTHrP and TNF that stimulate osteoclastic bone resorption. These factors enhance the expression of RANKL over OPG by osteoblasts, tipping the balance toward osteoclast activation thus causing bone resorption. This bone lysis stimulates the release of BMPs, TGFβ, IGFs and FGFs for stimulating the growth of metastatic cancer cells to bone. **Osteoblastic metastasis:** Factors released by osetoblastic cells, such as ET-1, Wnt, ERBB3, VEGF play an important role in osteoblastic metastasis by increasing cancer cell proliferation and enhance the effect of other growth factors including PDGF, FGFs, IGF-1. Osteoblast differentiation is also increased by BMPs through the activation of certain transcription factors. Urokinase Plasminogen Activator (uPA), a protease, also acts as mediator for osteoblastic bone metastasis by cleaving osteoclast-mediated bone resorption factors responsible for regulation of osteoclast differentiation; thereby blocking the bone resorption.

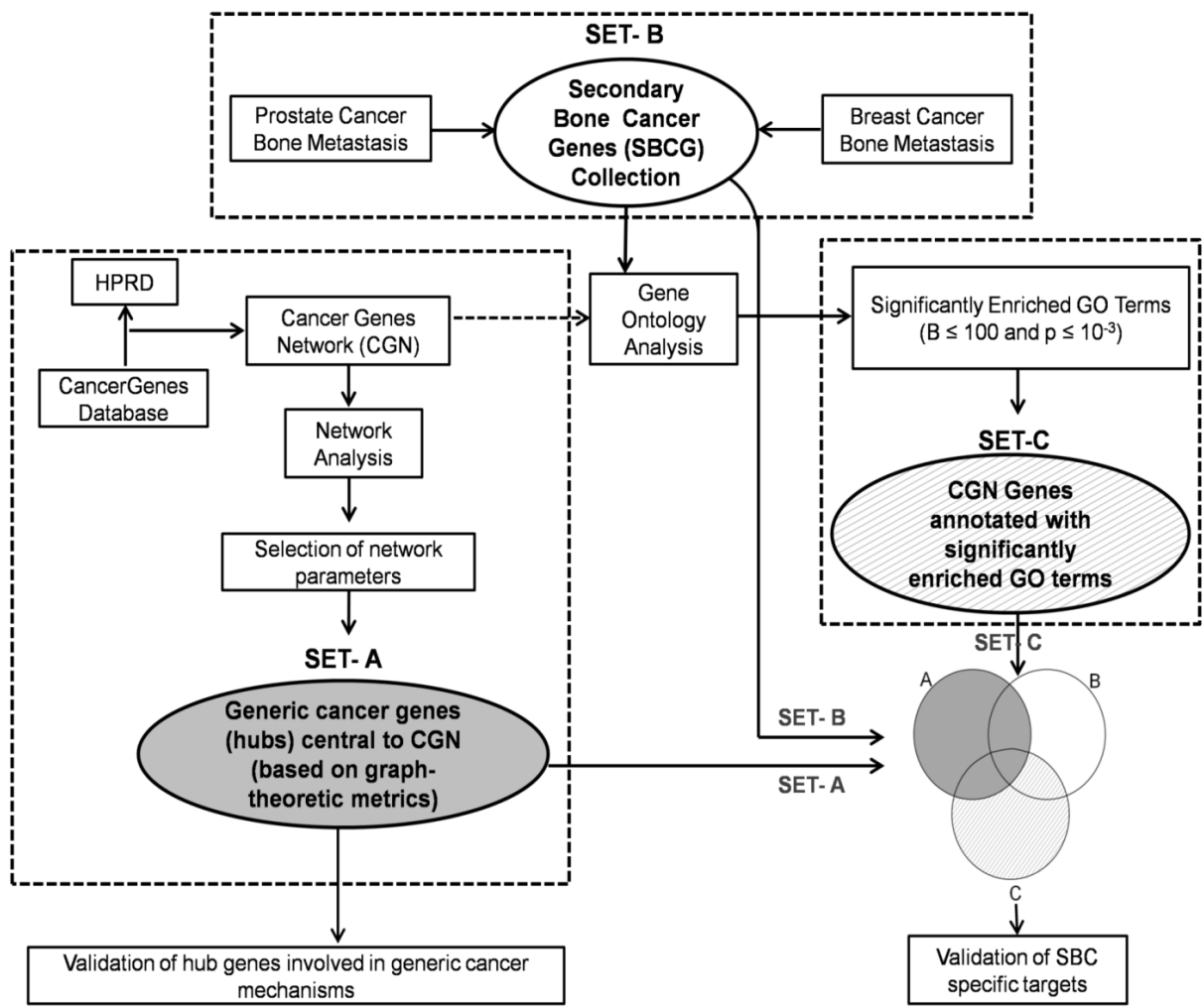

**Figure 2. Strategy implemented for the identification of targets specific to secondary bone cancer.**
The strategy that was implemented in this work comprised three tasks, leading to three corresponding gene-sets that were used for obtaining SBC-specific targets of potential therapeutic value. (a) A compilation of cancer genes from CancerGenes database was used to construct a representative cancer genes interactome (Cancer Genes Network; CGN) by mapping them on to a reference human protein interactome (Human Protein Reference Database; HPRD). Using methods of network analysis, proteins that are central to CGN and interaction dynamics were obtained. These set of genes (SET-A; shaded area) was found to be correlating well with genes implicated in generic cancer mechanisms (Figure 6) as well as those annotated as essential using mouse phenotype data (Figure 8). The CGN, comprising of 11602 interactions among 2665 proteins, also serves as a reference set (universe) for gene enrichment studies; (b) A set of genes (Secondary Bone Cancer Genes; SBCGs) that are implicated in metastasis to bone from primary breast and prostate cancer, the most prevalent causes of bone metastasis, was compiled from literature. This set (SET-B) serves as a basis of genes and ontological correlates of secondary bone cancer that characterize the disease phenotype; (c) Significantly enriched GO terms that characterize SBCGs were obtained by overrepresentation analysis against the 'cancer genes' universe. SET-C, a subset of CGN, was obtained by segregating cancer genes that were annotated with these SBC-specific ontological terms. Part of SET-C (hatched area; Set-c and Set-bc in Figure 10A) serves as a 'source set of target cancer genes' that, both, carry ontological essence of SBCGs and are not involved in generic cancer mechanisms. SBC-specific targets (Figure 3 and Figure 10B), that are annotated with key GO terms (Figure 9) reflecting role in, both, bone processes and metastasis mechanisms, were further obtained from the source set.

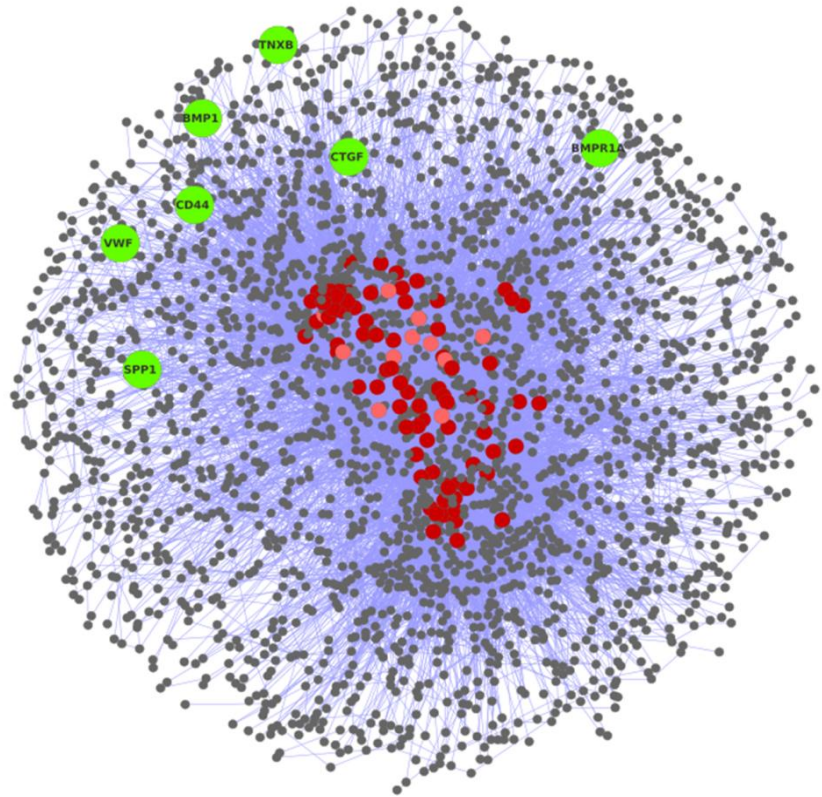

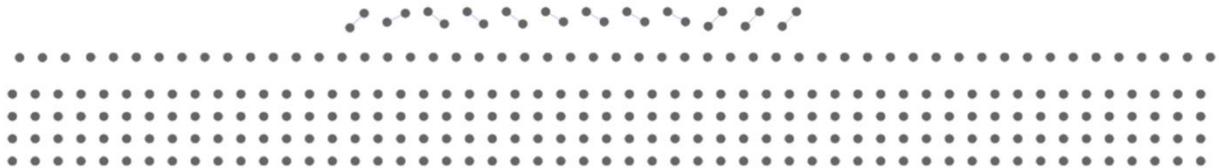

**Figure 3. CGN, a representative protein interactome of cancer genes, Top75 hub genes and SBC-specific targets.** Cancer Genes Network (CGN) is a protein interactome of cancer genes embodying molecular mechanisms of cancers. Each node represents a cancer protein and an edge between two nodes represents a protein-protein interaction. The giant cluster comprises 89% of all cancer genes. The hubs of CGN, cancer genes central to the structural stability and information dynamics, are depicted in shades of red. The hubs encode genes involved in generic cancer mechanisms and those classified as essential using phenotypic data from Mouse Genome Informatics. 88% of these hubs are essential ('dark red') and the rest non-essential ('light red'). SBC-specific targets, with ontological role in bone and metastasis processes, are highlighted in 'green'.

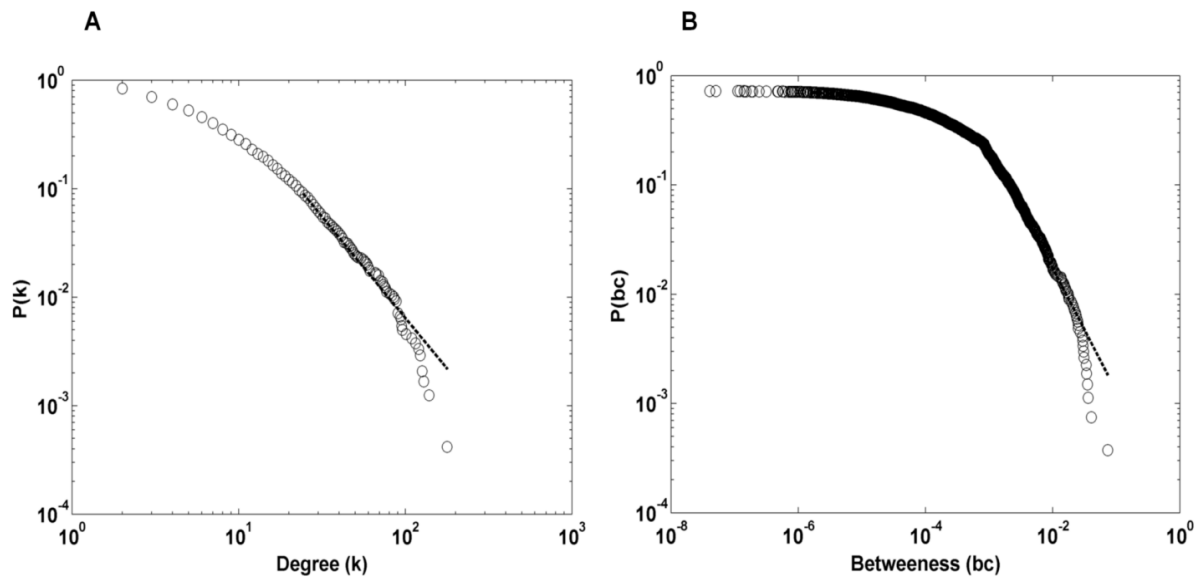

**Figure 4. Scale-free nature of degree and betweenness distributions of CGN.**
The distributions of (A) degree and (B) stress (as well as its normalized counterpart 'betweenness') show a scale-free nature. Dotted lines show the power law fit with an exponent of -2.85 and -2.13, respectively. This indicates presence of promiscuously interacting proteins with high degrees and central mediators that play role in information dynamics across the network, respectively.

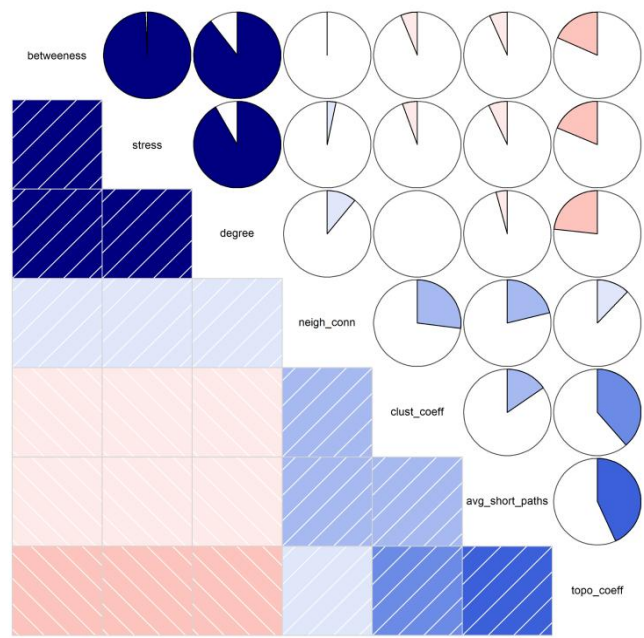

**Figure 5. Heatmap of topological metrics computed for Cancer Genes Network.**
The heatmap of pair-wise correlations among seven parameters that enumerate topological, dynamical and local clustering features of the network: betweenness, stress, degree, neighborhood connectivity (neigh_conn), clustering coefficient (clust_coeff), average shortest path length (avg_short_paths) and topological coefficient (topo_coeff). The heatmap highlights three parameters with very high mutual positive correlations (r=0.8989, 0.9930 and 0.9210): degree, betweenness and stress. The upper triangle of the heatmap depicts pair-wise correlations as pie charts. The lower triangle depicts positive and negative correlations in shades of blue and red, respectively; the darker the color the stronger the correlation. Positive and negative correlations are also depicted with right- and left-handed diagonal lines.

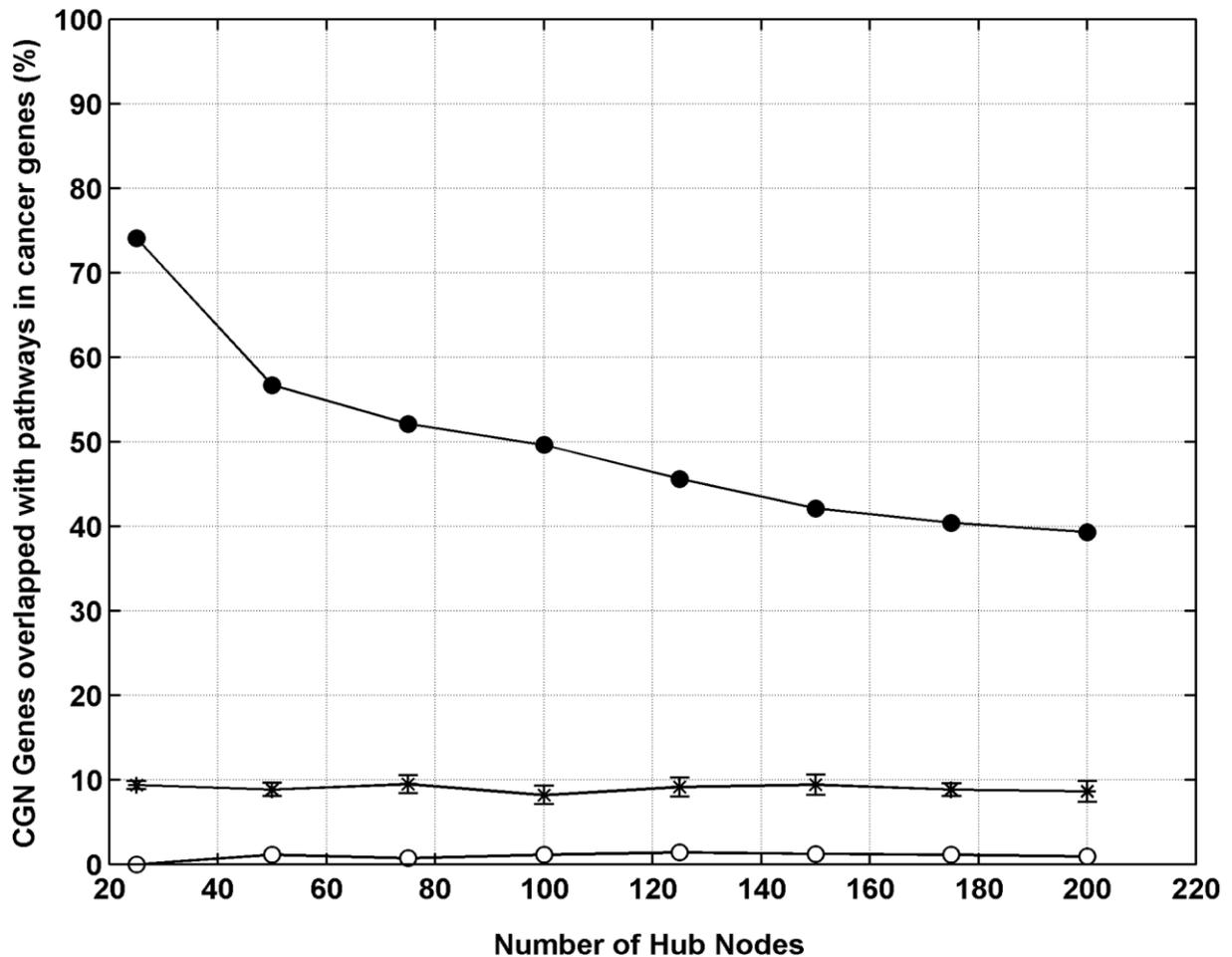

**Figure 6. Hubs of CGN correlate with generic cancer genes (KEGG-PIC).**
The hubs of CGN, identified using chosen network centrality parameters, correlate with the set of generic cancer genes (KEGG-PIC). For hub definitions varying between Top25—Top200, the CGN hubs (top ranked cancer genes) show good overlap with KEGG-PIC genes (filled circles). The correspondence of network centrality and generic role in cancers, expectedly, drops as the strictness of criterion used for identifying hubs is loosened. For the corresponding random samples, the overlap with KEGG-PIC is as expected (stars; error bars indicate standard errors from 1000 samples). The cancer genes with worst centrality rankings (open circles), show almost no overlap with generic cancer genes; worse than that expected from random samplings. Top75, Top50 and Top25 hubs, with 52%, 57% and 74% overlap with generic cancer genes, respectively, were further used for identification of SBC-specific cancer genes.

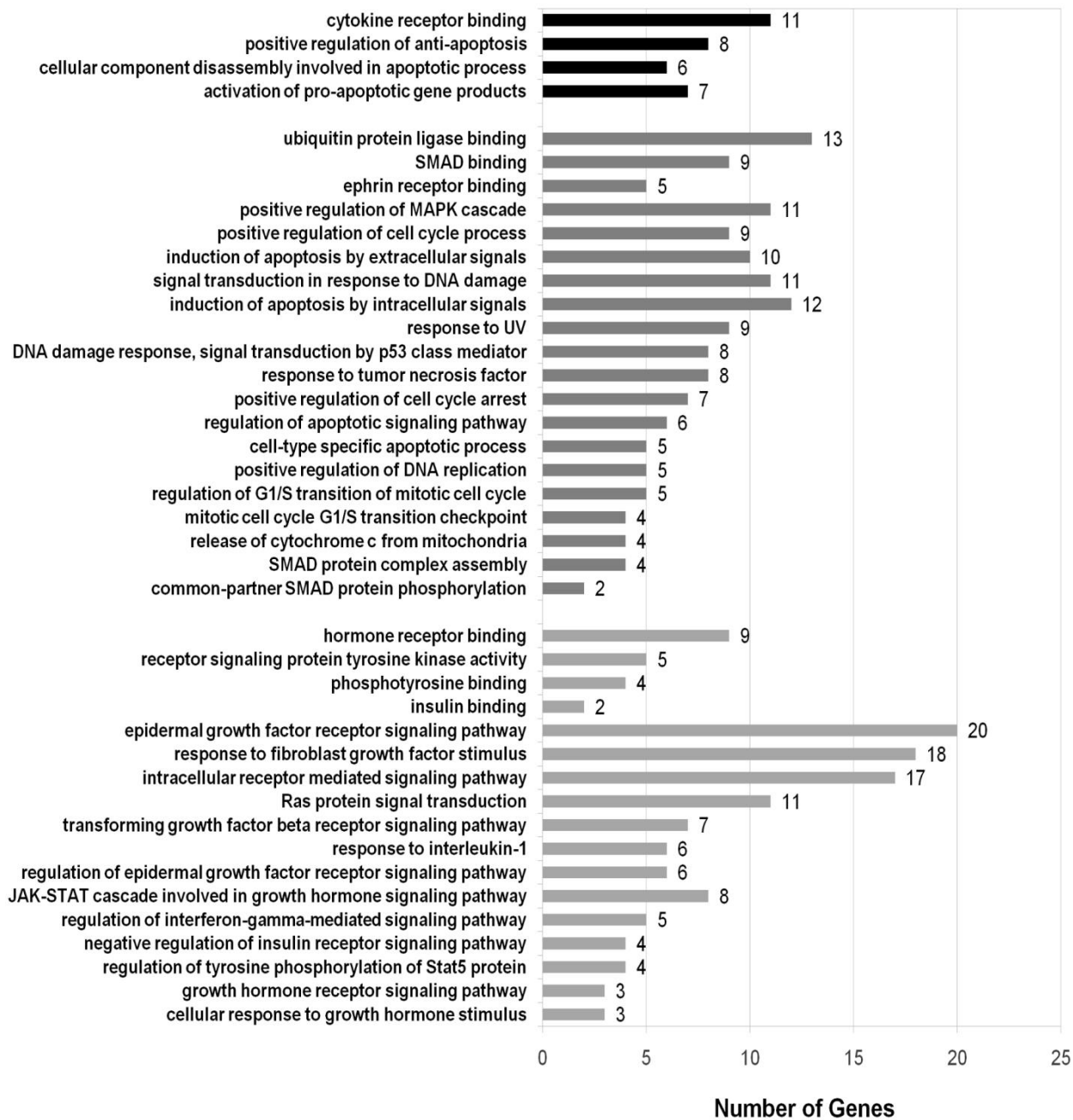

**Figure 7. Significantly enriched GO terms for hubs of CGN reflecting their role in generic cancer mechanisms.**
The overrepresentation studies of Top75 CGN hubs reflect their role in generic cancer processes and functions. Among the significantly enriched GO terms, 17 represent self-proliferation circuits (light-gray), 20 represent cytostasis and differentiation circuits (dark gray) and 4 represent viability circuits (black).

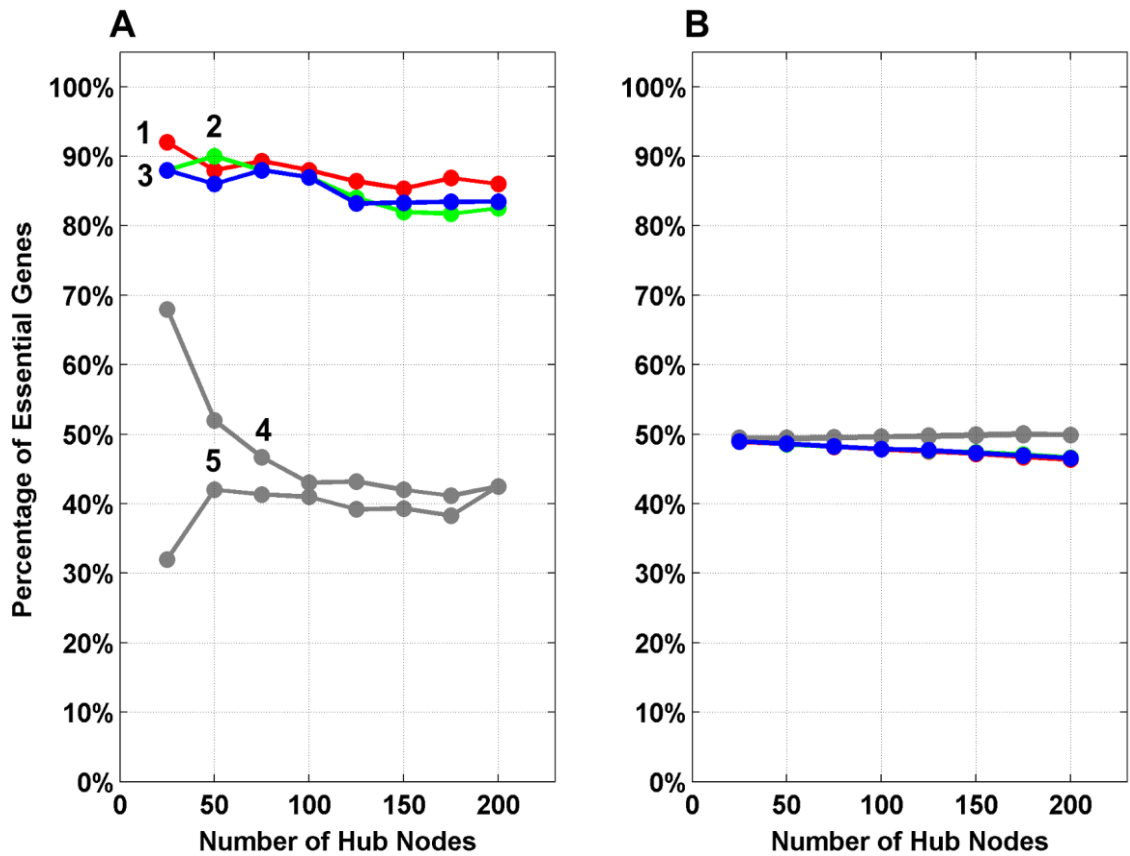

**Figure 8. The percentage of essential genes in the hubs and the corresponding non-hubs of CGN.**
A. For hub definitions varying between Top25 to Top200, the hubs identified using degree (1, 'red'), betweenness (2, 'green'), and stress (3, 'blue') have significantly high percentage (82%—92%) of essential genes, classified using phenotypic data for Mouse Genome Informatics. Among the rest of the four parameters neighborhood connectivity (4) and topological coefficient (5) show neither significant nor consistent correlation with essential genes. Due to the nature of 'clustering coefficient' and 'average shortest path length' parameters, the data could not be binned at the same intervals. For clustering coefficient, percentage of essential genes among the nodes having up to Top200 rankings is in the range of 27% and 62%. For average shortest path length, the percentage of essential genes for nodes having ranking up to Top200 is 26%, worse than expected from random sampling. B. In the corresponding non-hubs, the percentage of essential genes is as expected from random sampling, regardless of the centrality measure or cut-off used for hub definition.

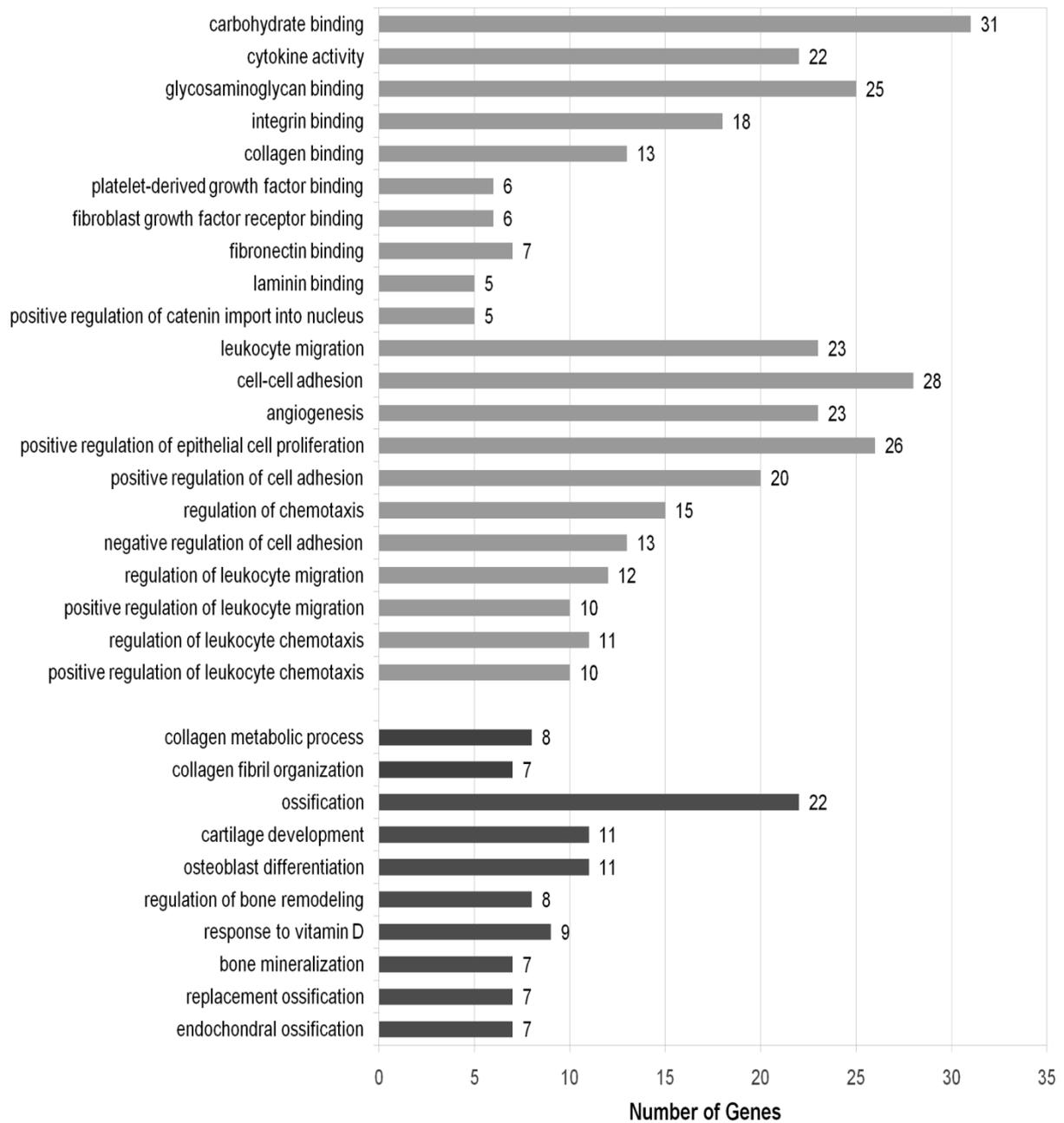

**Figure 9. Overrepresented GO terms of processes relevant and necessary for execution of bone metastasis.**
From the GO enrichment studies of SBCGs, curated from literature, 21 GO terms relevant for metastasis ('light gray') and 10 GO terms relevant for bone ('dark gray') processes were identified.

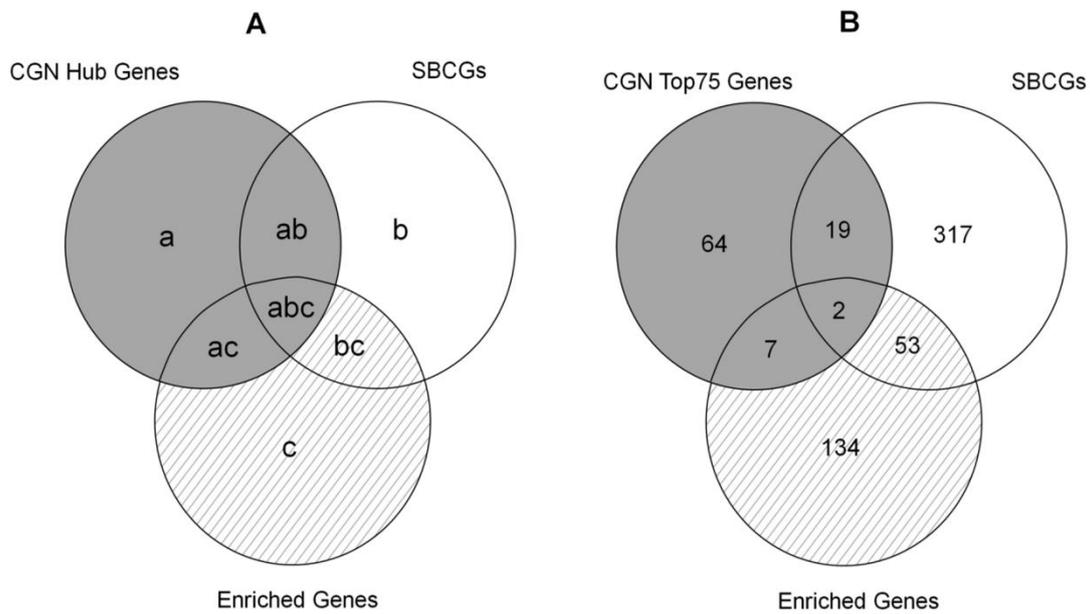

**Figure 10. Venn diagrams depicting the strategy used for the identification of SBC-specific targets.**
A. Venn diagram legend representing generic cancer genes identified by network analysis (CGN hub genes; gray area) and gene subsets used as a source of target genes that are specific to bone metastasis (hatched area). In search of SBC-specific targets, the generic cancer genes were rejected and the potential source of target genes was refined to obtain the final targets. B. Venn diagram depicting components of gene-sets identified using Top75 hubs. The source set of target genes (53+134) was refined to obtain seven targets.

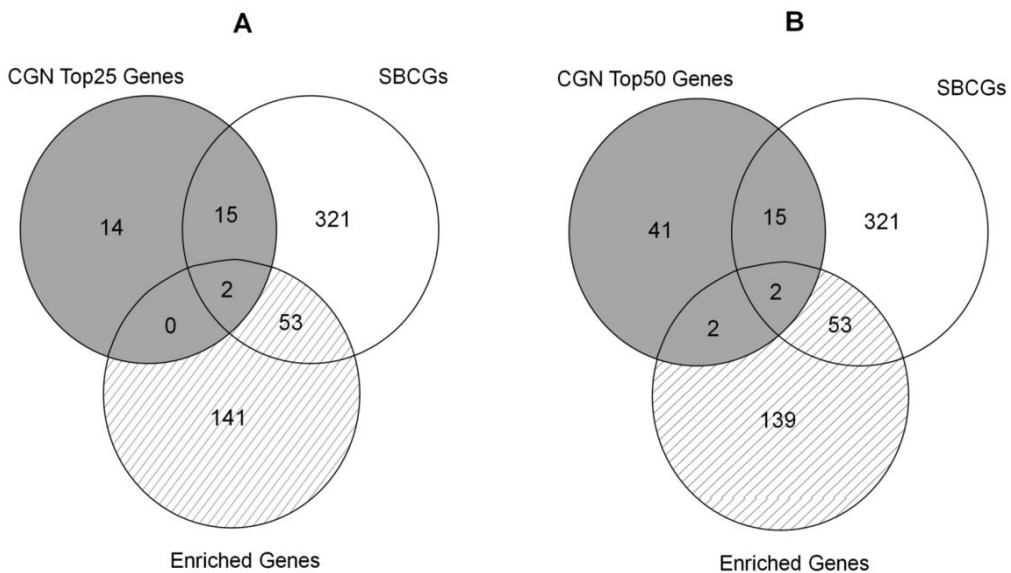

**Figure 11. Venn diagram of gene sets obtained with Top25 and Top50 CGN hubs.** The Venn diagrams with generic gene-set (gray area) and source set of target genes (hatched area) for (A) Top25 and (B) Top50 CGN hubs.

# SUPPLEMENTARY INFORMATION S1

An approach for the identification of targets specific to bone metastasis using cancer genes interactome and gene ontology analysis

Shikha Vashisht and Ganesh Bagler*

Biotechnology Division, CSIR-IHBT, Institute of Himalayan Bioresource Technology, Council of Scientific and Industrial Research, Palampur, 176061 (H.P.), India

## SUMMARY

In the supplementary information we present the details of hubs of CGN; the KEGG-PIC genes that serve as a reference set of generic cancer genes; secondary bone cancer genes that were curated and compiled; characteristic GO terms that were used to obtain SBC-specific targets and relevance of SBC-specific targets identified from experimentally validated studies. The supplementary information contains 5 tables, out of which Table S1.3 has 46 references, Table S1.4 has 23 references and Table S1.5 has 8 references.

## CONTENTS

**Table S1.1.** Hub genes (Top25—Top200) of Cancer Genes Network.

**Table S1.2.** Details of KEGG-PIC genes (328) from KEGG PATHWAY Database.

**Table S1.3.** SBCGs (391) compiled for metastasis of primary breast and prostate cancer to bone.

**Table S1.4.** Significantly enriched GO terms, characteristic to metastasis to bone, identified from enrichment analysis of SBCGs.

**Table S1.5.** Relevance of SBC targets.

**Table S1.1. Hub genes (Top25—Top200) of Cancer Genes Network.**

| Top25 hubs (31) | Top50 hubs (60) | Top75 hubs (92) | Top100 hubs (125) | Top125 hubs (158) | Top150 hubs (185) | Top175 hubs (213) | Top200 hubs (247) |
|---|---|---|---|---|---|---|---|
| ABL1 | ABL1 | ABL1 | ABL1 | ABL1 | ABL1 | ABL1 | ABL1 |
| AKT1 | AKT1 | ACTB | ACTB | ACTB | ACTB | ACTB | ACTB |
| AR | AR | AKT1 | AKT1 | AKT1 | AKT1 | ACVR1 | ACVR1 |
| BRCA1 | BCL2 | APP | APP | APP | APP | AKT1 | AKT1 |
| CASP3 | BRCA1 | AR | AR | AR | AR | APP | APP |
| CREBBP | CASP3 | BCL2 | BCL2 | AXIN1 | ATM | AR | AR |
| CSNK2A1 | CAV1 | BRCA1 | BRCA1 | BCL2 | AXIN1 | ATM | ATF2 |
| CTNNB1 | CBL | CASP3 | CASP3 | BRCA1 | BAT3 | AXIN1 | ATM |
| EGFR | CDK2 | CASP8 | CASP8 | BTK | BCL2 | BAT3 | AXIN1 |
| EP300 | CREBBP | CAV1 | CAV1 | CALR | BCR | BCAR1 | BAD |
| ESR1 | CSNK2A1 | CBL | CBL | CASP3 | BRCA1 | BCL2 | BAT3 |
| FYN | CTNNB1 | CDK2 | CDK2 | CASP8 | BTK | BCR | BCAR1 |
| GRB2 | DLG4 | CHUK | CDKN1A | CAV1 | CADM1 | BIRC2 | BCL2 |
| HDAC1 | EGFR | CREBBP | CDKN1B | CBL | CALR | BRCA1 | BCR |
| JUN | EP300 | CRK | CEBPB | CCNB1 | CASP3 | BRCA2 | BIRC2 |
| MAPK1 | ESR1 | CSNK2A1 | CHUK | CCND1 | CASP8 | BTK | BMPR1B |
| MAPK3 | FLNA | CTNNB1 | CREBBP | CDC42 | CAV1 | CADM1 | BRCA1 |
| PRKACA | FYN | DLG4 | CRK | CDK2 | CBL | CALR | BRCA2 |
| PRKCA | GRB2 | EEF1A1 | CSNK2A1 | CDK9 | CCNB1 | CASP3 | BTK |
| RAF1 | GSK3B | EGFR | CSNK2A2 | CDKN1A | CCND1 | CASP8 | BTRC |
| RB1 | HDAC1 | EP300 | CTNNB1 | CDKN1B | CDC42 | CAV1 | CADM1 |
| RELA | HSP90AA1 | ESR1 | DLG4 | CEBPB | CDK2 | CBL | CALR |
| SMAD2 | INSR | EWSR1 | DYNLL1 | CHUK | CDK4 | CCNB1 | CASP3 |
| SMAD3 | JAK2 | FLNA | EEF1A1 | CREBBP | CDK9 | CCND1 | CASP8 |
| SMAD4 | JUN | FYN | EGFR | CRK | CDKN1A | CDC42 | CAV1 |
| SRC | LCK | GNAI2 | EP300 | CRKL | CDKN1B | CDK2 | CBL |
| STAT3 | LYN | GRB2 | EPHB2 | CSNK2A1 | CEBPB | CDK4 | CCNB1 |
| TGFBR1 | MAPK1 | GSK3B | ERBB2 | CSNK2A2 | CHUK | CDK9 | CCND1 |
| TP53 | MAPK14 | HDAC1 | ESR1 | CTNNB1 | CREB1 | CDKN1A | CD44 |
| YWHAG | MAPK3 | HRAS | EWSR1 | DAXX | CREBBP | CDKN1B | CDC42 |
| YWHAZ | NFKB1 | HSP90AA1 | FLNA | DLG4 | CRK | CEBPB | CDH1 |
|  | NR3C1 | IGF1R | FOS | DYNLL1 | CRKL | CHUK | CDK2 |
|  | PAK1 | INSR | FYN | EEF1A1 | CSK | CREB1 | CDK4 |
|  | PCNA | IRS1 | GNAI2 | EGFR | CSNK2A1 | CREBBP | CDK9 |
|  | PLCG1 | JAK1 | GRB2 | EP300 | CSNK2A2 | CRK | CDKN1A |
|  | PRKACA | JAK2 | GSK3B | EPB41L3 | CTNNB1 | CRKL | CDKN1B |
|  | PRKCA | JUN | HDAC1 | EPHB2 | DAXX | CSK | CDKN2A |
|  | PRKCD | KAT5 | HDAC2 | ERBB2 | DLG4 | CSNK2A1 | CEBPB |
|  | PTK2 | KIT | HIF1A | ESR1 | DYNLL1 | CSNK2A2 | CHD3 |
|  | PTPN11 | LCK | HMGB1 | EWSR1 | E2F1 | CTNNB1 | CHUK |
|  | RAF1 | LRP1 | HRAS | FLNA | EEF1A1 | DAXX | CREB1 |
|  | RB1 | LYN | HSP90AA1 | FN1 | EGFR | DLG4 | CREBBP |
|  | RELA | MAPK1 | HTT | FOS | EP300 | DYNLL1 | CRK |

| | | | | | | |
|---|---|---|---|---|---|---|
| SMAD2 | MAPK14 | IGF1R | FYN | EPB41L3 | E2F1 | CRKL |
| SMAD3 | MAPK3 | IKBKB | GNAI2 | EPHB2 | EEF1A1 | CSK |
| SMAD4 | MAPK8 | IKBKG | GNB2L1 | ERBB2 | EGFR | CSNK2A1 |
| SP1 | MDM2 | INSR | GRB2 | ESR1 | EP300 | CSNK2A2 |
| SRC | MYC | IRS1 | GSK3B | ESR2 | EPB41L3 | CSNK2B |
| STAT1 | NCK1 | JAK1 | HCK | EWSR1 | EPHB2 | CTNNB1 |
| STAT3 | NFKB1 | JAK2 | HDAC1 | FLNA | EPOR | DAXX |
| TGFBR1 | NFKBIA | JUN | HDAC2 | FN1 | ERBB2 | DLG4 |
| TP53 | NOTCH1 | KAT5 | HDAC3 | FOS | ESR1 | DOK1 |
| TRAF2 | NR3C1 | KIT | HIF1A | FYN | ESR2 | DVL1 |
| TRAF6 | PAK1 | LCK | HIPK2 | GNAI2 | ETS1 | DYNLL1 |
| UBB | PCNA | LRP1 | HMGB1 | GNB2L1 | EWSR1 | E2F1 |
| UBE2I | PLCG1 | LYN | HRAS | GRB2 | FGFR1 | EEF1A1 |
| VIM | PML | MAPK1 | HSP90AA1 | GSK3B | FLNA | EGFR |
| YWHAB | PRKACA | MAPK14 | HSPA1A | HCK | FN1 | EP300 |
| YWHAG | PRKCA | MAPK3 | HTT | HDAC1 | FOS | EPB41 |
| YWHAZ | PRKCD | MAPK8 | IGF1R | HDAC2 | FYN | EPB41L3 |
| | PRKCZ | MDM2 | IKBKB | HDAC3 | GNA13 | EPHB2 |
| | PTK2 | MYC | IKBKG | HIF1A | GNAI2 | EPOR |
| | PTK2B | NCK1 | INSR | HIPK2 | GNB2L1 | ERBB2 |
| | PTPN11 | NCOA1 | IRS1 | HMGB1 | GRB2 | ERBB4 |
| | PTPN6 | NCOR2 | JAK1 | HRAS | GSK3B | ESR1 |
| | RAC1 | NFKB1 | JAK2 | HSP90AA1 | HCK | ESR2 |
| | RAF1 | NFKBIA | JUN | HSPA1A | HDAC1 | ETS1 |
| | RASA1 | NFKBIB | KAT2B | HTT | HDAC2 | EWSR1 |
| | RB1 | NOTCH1 | KAT5 | IGF1R | HDAC3 | FADD |
| | RELA | NR3C1 | KIT | IKBKB | HIF1A | FGFR1 |
| | SMAD1 | PAK1 | LCK | IKBKG | HIPK2 | FLNA |
| | SMAD2 | PCNA | LRP1 | INSR | HMGB1 | FN1 |
| | SMAD3 | PDGFRB | LYN | IRS1 | HRAS | FOS |
| | SMAD4 | PIN1 | MAP3K7 | ITGB1 | HSP90AA1 | FYN |
| | SP1 | PLCG1 | MAPK1 | JAK1 | HSPA1A | GNA13 |
| | SRC | PML | MAPK14 | JAK2 | HTT | GNAI2 |
| | STAT1 | POLR2A | MAPK3 | JUN | IGF1R | GNB2L1 |
| | STAT3 | PRKACA | MAPK8 | KAT2B | IGF2 | GRB2 |
| | SUMO4 | PRKCA | MBP | KAT5 | IKBKB | GSK3B |
| | SYK | PRKCD | MDM2 | KDR | IKBKG | HCK |
| | TGFBR1 | PRKCZ | MET | KIT | INSR | HDAC1 |
| | TP53 | PTCH1 | MYC | LCK | IRS1 | HDAC2 |
| | TRAF2 | PTEN | MYOD1 | LRP1 | ITGB1 | HDAC3 |
| | TRAF6 | PTK2 | NCK1 | LYN | ITGB4 | HGS |
| | UBB | PTK2B | NCOA1 | MAP3K14 | JAK1 | HIF1A |
| | UBE2I | PTPN11 | NCOA3 | MAP3K7 | JAK2 | HIPK2 |
| | VAV1 | PTPN6 | NCOR1 | MAPK1 | JUN | HMGB1 |
| | VIM | RAC1 | NCOR2 | MAPK14 | KAT2B | HRAS |
| | YWHAB | RAD51 | NFKB1 | MAPK3 | KAT5 | HSP90AA1 |
| | YWHAG | RAF1 | NFKBIA | MAPK8 | KDR | HSPA1A |
| | | YWHAZ | RASA1 | NFKBIB | MAPT | KHDRBS1 | HTT |
| | | ZBTB16 | RB1 | NOTCH1 | MBP | KIT | IGF1R |
| | | | RELA | NR3C1 | MDM2 | KPNB1 | IGF2 |
| | | | RXRA | PAK1 | MET | KRT18 | IKBKB |
| | | | SIN3A | PCNA | MYC | LCK | IKBKG |
| | | | SMAD1 | PDGFRB | MYOD1 | LRP1 | IL6ST |
| | | | SMAD2 | PIAS1 | NCK1 | LRP6 | INSR |
| | | | SMAD3 | PIN1 | NCOA1 | LYN | IRAK1 |
| | | | SMAD4 | PLCG1 | NCOA3 | MAP3K14 | IRS1 |
| | | | SMURF2 | PML | NCOR1 | MAP3K5 | IRS2 |
| | | | SP1 | POLR2A | NCOR2 | MAP3K7 | ITGB1 |
| | | | SRC | PPP2CA | NDRG1 | MAPK1 | ITGB3 |
| | | | STAT1 | PRKACA | NFKB1 | MAPK14 | ITGB4 |
| | | | STAT3 | PRKAR2A | NFKBIA | MAPK3 | JAK1 |
| | | | STAT5A | PRKCA | NFKBIB | MAPK8 | JAK2 |
| | | | SUMO1 | PRKCD | NOTCH1 | MAPK9 | JUN |
| | | | SUMO4 | PRKCZ | NR3C1 | MAPT | KAT2B |
| | | | SVIL | PTCH1 | PAK1 | MBP | KAT5 |
| | | | SYK | PTEN | PCNA | MDM2 | KDR |
| | | | TBP | PTK2 | PDGFRB | MET | KHDRBS1 |
| | | | TGFBR1 | PTK2B | PIAS1 | MMP2 | KIT |
| | | | TNFRSF1A | PTPN11 | PIN1 | MYC | KPNB1 |
| | | | TP53 | PTPN6 | PLCG1 | MYOD1 | KRT18 |
| | | | TRAF2 | PTPRC | PLSCR1 | NCK1 | LCK |
| | | | TRAF6 | RAC1 | PML | NCOA1 | LRP1 |
| | | | UBB | RAD51 | POLR2A | NCOA2 | LRP6 |
| | | | UBE2I | RAF1 | PPP2CA | NCOA3 | LYN |
| | | | VAV1 | RASA1 | PRKAA1 | NCOR1 | MAP2K1 |
| | | | VIM | RB1 | PRKACA | NCOR2 | MAP3K14 |
| | | | XRCC6 | RELA | PRKAR2A | NDRG1 | MAP3K5 |
| | | | YWHAB | RET | PRKCA | NFKB1 | MAP3K7 |
| | | | YWHAG | RHOA | PRKCD | NFKBIA | MAPK1 |
| | | | YWHAH | RXRA | PRKCZ | NFKBIB | MAPK14 |
| | | | YWHAZ | SIN3A | PRKDC | NOTCH1 | MAPK3 |
| | | | ZBTB16 | SMAD1 | PSEN1 | NR3C1 | MAPK8 |
| | | | | SMAD2 | PTCH1 | PAK1 | MAPK9 |
| | | | | SMAD3 | PTEN | PCNA | MAPT |
| | | | | SMAD4 | PTK2 | PDGFRB | MBP |
| | | | | SMURF2 | PTK2B | PDPK1 | MDM2 |
| | | | | SP1 | PTPN1 | PIAS1 | MET |
| | | | | SRC | PTPN11 | PIN1 | MITF |
| | | | | STAT1 | PTPN6 | PLCG1 | MLLT4 |
| | | | | STAT3 | PTPRC | PLSCR1 | MMP2 |
| | | | | STAT5A | RAC1 | PML | MMP9 |
| | | | | STAT5B | RAD51 | POLR2A | MPP3 |
| | | | | SUMO1 | RAF1 | PPP2CA | MYC |
| | | | | SUMO4 | RARA | PPP2R1B | MYOD1 |
| | | | | SVIL | RASA1 | PRKAA1 | NCK1 |

| | | | | | | |
|---|---|---|---|---|---|---|
| | | | SYK | RB1 | PRKACA | NCOA1 |
| | | | TBP | RELA | PRKAR2A | NCOA2 |
| | | | TGFBR1 | RET | PRKCA | NCOA3 |
| | | | TNFRSF1A | RHOA | PRKCD | NCOA6 |
| | | | TP53 | RXRA | PRKCZ | NCOR1 |
| | | | TRAF2 | SIN3A | PRKDC | NCOR2 |
| | | | TRAF6 | SKP1 | PSEN1 | NDRG1 |
| | | | TUBB | SMAD1 | PTCH1 | NFKB1 |
| | | | UBB | SMAD2 | PTEN | NFKBIA |
| | | | UBE2I | SMAD3 | PTK2 | NFKBIB |
| | | | VAV1 | SMAD4 | PTK2B | NGFR |
| | | | VIM | SMURF2 | PTPN1 | NOTCH1 |
| | | | XRCC6 | SOCS1 | PTPN11 | NR3C1 |
| | | | YWHAB | SOS1 | PTPN6 | PAK1 |
| | | | YWHAE | SP1 | PTPRC | PARP1 |
| | | | YWHAG | SRC | RAC1 | PCNA |
| | | | YWHAH | STAT1 | RAD51 | PDGFRB |
| | | | YWHAQ | STAT3 | RAF1 | PDPK1 |
| | | | YWHAZ | STAT5A | RARA | PIAS1 |
| | | | ZBTB16 | STAT5B | RASA1 | PIK3R2 |
| | | | | SUMO1 | RB1 | PIN1 |
| | | | | SUMO4 | RELA | PLCG1 |
| | | | | SVIL | RET | PLSCR1 |
| | | | | SYK | RHOA | PML |
| | | | | TBP | RUNX1 | POLR2A |
| | | | | TGFBR1 | RXRA | PPP2CA |
| | | | | TNFRSF1A | SIN3A | PPP2R1B |
| | | | | TP53 | SKIL | PRKAA1 |
| | | | | TP73 | SKP1 | PRKACA |
| | | | | TRAF2 | SMAD1 | PRKAR2A |
| | | | | TRAF6 | SMAD2 | PRKCA |
| | | | | TSC2 | SMAD3 | PRKCD |
| | | | | TUBB | SMAD4 | PRKCZ |
| | | | | UBB | SMAD7 | PRKDC |
| | | | | UBE2I | SMARCA4 | PSEN1 |
| | | | | VAV1 | SMURF1 | PTCH1 |
| | | | | VIM | SMURF2 | PTEN |
| | | | | XPO1 | SNCA | PTK2 |
| | | | | XRCC6 | SOCS1 | PTK2B |
| | | | | YWHAB | SOS1 | PTPN1 |
| | | | | YWHAE | SP1 | PTPN11 |
| | | | | YWHAG | SRC | PTPN6 |
| | | | | YWHAH | STAT1 | PTPRC |
| | | | | YWHAQ | STAT3 | RAC1 |
| | | | | YWHAZ | STAT5A | RAD51 |
| | | | | ZAP70 | STAT5B | RAF1 |
| | | | | ZBTB16 | SUMO1 | RARA |
| | | | | | SUMO4 | RASA1 |

| | | | | | | | | | |
|---|---|---|---|---|---|---|---|---|---|
| | | | | | | | | SVIL | RB1 |
| | | | | | | | | SYK | RBL1 |
| | | | | | | | | TBP | RELA |
| | | | | | | | | TCF3 | RELB |
| | | | | | | | | TGFBR1 | RET |
| | | | | | | | | TNFRSF1A | RHOA |
| | | | | | | | | TP53 | RIPK1 |
| | | | | | | | | TP73 | RUNX1 |
| | | | | | | | | TRAF2 | RXRA |
| | | | | | | | | TRAF6 | SIN3A |
| | | | | | | | | TSC2 | SKIL |
| | | | | | | | | TUBB | SKP1 |
| | | | | | | | | UBB | SKP2 |
| | | | | | | | | UBE2I | SMAD1 |
| | | | | | | | | VAV1 | SMAD2 |
| | | | | | | | | VHL | SMAD3 |
| | | | | | | | | VIM | SMAD4 |
| | | | | | | | | XPO1 | SMAD7 |
| | | | | | | | | XRCC6 | SMARCA4 |
| | | | | | | | | YWHAB | SMURF1 |
| | | | | | | | | YWHAE | SMURF2 |
| | | | | | | | | YWHAG | SNCA |
| | | | | | | | | YWHAH | SOCS1 |
| | | | | | | | | YWHAQ | SOS1 |
| | | | | | | | | YWHAZ | SP1 |
| | | | | | | | | ZAP70 | SRC |
| | | | | | | | | ZBTB16 | STAT1 |
| | | | | | | | | | STAT3 |
| | | | | | | | | | STAT5A |
| | | | | | | | | | STAT5B |
| | | | | | | | | | SUMO1 |
| | | | | | | | | | SUMO4 |
| | | | | | | | | | SVIL |
| | | | | | | | | | SYK |
| | | | | | | | | | TBP |
| | | | | | | | | | TCF3 |
| | | | | | | | | | TERT |
| | | | | | | | | | TGFBR1 |
| | | | | | | | | | TGFBR2 |
| | | | | | | | | | TNFRSF1A |
| | | | | | | | | | TP53 |
| | | | | | | | | | TP73 |
| | | | | | | | | | TRAF2 |
| | | | | | | | | | TRAF6 |
| | | | | | | | | | TSC2 |
| | | | | | | | | | TUBB |

| | | | | | | | | | UBB |
| --- | --- | --- | --- | --- | --- | --- | --- | --- | --- |
| | | | | | | | | | UBE2I |
| | | | | | | | | | VAV1 |
| | | | | | | | | | VHL |
| | | | | | | | | | VIM |
| | | | | | | | | | XPO1 |
| | | | | | | | | | XRCC6 |
| | | | | | | | | | YWHAB |
| | | | | | | | | | YWHAE |
| | | | | | | | | | YWHAG |
| | | | | | | | | | YWHAH |
| | | | | | | | | | YWHAQ |
| | | | | | | | | | YWHAZ |
| | | | | | | | | | ZAP70 |
| | | | | | | | | | ZBTB16 |

**Table S1.2. Details of KEGG-PIC genes (328) from KEGG PATHWAY Database.**

| HGNC Gene ID | Gene Symbol | Full Name of the Gene |
|---|---|---|
| 1630 | DCC | deleted in colorectal carcinoma |
| 836 | CASP3 | caspase 3, apoptosis-related cysteine peptidase (EC:3.4.22.56) |
| 842 | CASP9 | caspase 9, apoptosis-related cysteine peptidase (EC:3.4.22.62) |
| 26060 | APPL1 | adaptor protein, phosphotyrosine interaction, PH domain and leucine zipper containing 1 |
| 999 | CDH1 | cadherin 1, type 1, E-cadherin (epithelial) |
| 1499 | CTNNB1 | catenin (cadherin-associated protein), beta 1, 88kDa |
| 1496 | CTNNA2 | catenin (cadherin-associated protein), alpha 2 |
| 1495 | CTNNA1 | catenin (cadherin-associated protein), alpha 1, 102kDa |
| 29119 | CTNNA3 | catenin (cadherin-associated protein), alpha 3 |
| 8312 | AXIN1 | axin 1 |
| 8313 | AXIN2 | axin 2 |
| 10297 | APC2 | adenomatosis polyposis coli 2 |
| 324 | APC | adenomatous polyposis coli |
| 2932 | GSK3B | glycogen synthase kinase 3 beta (EC:2.7.11.1 2.7.11.26) |
| 6932 | TCF7 | transcription factor 7 (T-cell specific, HMG-box) |
| 83439 | TCF7L1 | transcription factor 7-like 1 (T-cell specific, HMG-box) |
| 6934 | TCF7L2 | transcription factor 7-like 2 (T-cell specific, HMG-box) |
| 51176 | LEF1 | lymphoid enhancer-binding factor 1 |
| 332 | BIRC5 | baculoviral IAP repeat containing 5 |
| 4609 | MYC | v-myc myelocytomatosis viral oncogene homolog (avian) |
| 595 | CCND1 | cyclin D1 |
| 7471 | WNT1 | wingless-type MMTV integration site family, member 1 |
| 7482 | WNT2B | wingless-type MMTV integration site family, member 2B |
| 7472 | WNT2 | wingless-type MMTV integration site family member 2 |
| 7473 | WNT3 | wingless-type MMTV integration site family, member 3 |
| 89780 | WNT3A | wingless-type MMTV integration site family, member 3A |
| 54361 | WNT4 | wingless-type MMTV integration site family, member 4 |
| 7474 | WNT5A | wingless-type MMTV integration site family, member 5A |
| 81029 | WNT5B | wingless-type MMTV integration site family, member 5B |
| 7475 | WNT6 | wingless-type MMTV integration site family, member 6 |
| 7477 | WNT7B | wingless-type MMTV integration site family, member 7B |
| 7476 | WNT7A | wingless-type MMTV integration site family, member 7A |
| 7479 | WNT8B | wingless-type MMTV integration site family, member 8B |
| 7478 | WNT8A | wingless-type MMTV integration site family, member 8A |
| 7483 | WNT9A | wingless-type MMTV integration site family, member 9A |
| 7484 | WNT9B | wingless-type MMTV integration site family, member 9B |
| 7480 | WNT10B | wingless-type MMTV integration site family, member 10B |
| 80326 | WNT10A | wingless-type MMTV integration site family, member 10A |
| 7481 | WNT11 | wingless-type MMTV integration site family, member 11 |

| 51384 | WNT16 | wingless-type MMTV integration site family, member 16 |
|---|---|---|
| 8321 | FZD1 | frizzled family receptor 1 |
| 8324 | FZD7 | frizzled family receptor 7 |
| 2535 | FZD2 | frizzled family receptor 2 |
| 7976 | FZD3 | frizzled family receptor 3 |
| 8322 | FZD4 | frizzled family receptor 4 |
| 8325 | FZD8 | frizzled family receptor 8 |
| 7855 | FZD5 | frizzled family receptor 5 |
| 8323 | FZD6 | frizzled family receptor 6 |
| 8326 | FZD9 | frizzled family receptor 9 |
| 11211 | FZD10 | frizzled family receptor 10 |
| 1856 | DVL2 | dishevelled, dsh homolog 2 (Drosophila) |
| 1857 | DVL3 | dishevelled, dsh homolog 3 (Drosophila) |
| 1855 | DVL1 | dishevelled, dsh homolog 1 (Drosophila) |
| 1284 | COL4A2 | collagen, type IV, alpha 2 |
| 1286 | COL4A4 | collagen, type IV, alpha 4 |
| 1288 | COL4A6 | collagen, type IV, alpha 6 |
| 1287 | COL4A5 | collagen, type IV, alpha 5 |
| 1282 | COL4A1 | collagen, type IV, alpha 1 |
| 284217 | LAMA1 | laminin, alpha 1 |
| 3908 | LAMA2 | laminin, alpha 2 |
| 3911 | LAMA5 | laminin, alpha 5 |
| 3909 | LAMA3 | laminin, alpha 3 |
| 3910 | LAMA4 | laminin, alpha 4 |
| 3912 | LAMB1 | laminin, beta 1 |
| 3913 | LAMB2 | laminin, beta 2 (laminin S) |
| 3914 | LAMB3 | laminin, beta 3 |
| 22798 | LAMB4 | laminin, beta 4 |
| 3915 | LAMC1 | laminin, gamma 1 (formerly LAMB2) |
| 3918 | LAMC2 | laminin, gamma 2 |
| 10319 | LAMC3 | laminin, gamma 3 |
| 2335 | FN1 | fibronectin 1 |
| 3673 | ITGA2 | integrin, alpha 2 (CD49B, alpha 2 subunit of VLA-2 receptor) |
| 3674 | ITGA2B | integrin, alpha 2b (platelet glycoprotein IIb of IIb/IIIa complex, antigen CD41) |
| 3675 | ITGA3 | integrin, alpha 3 (antigen CD49C, alpha 3 subunit of VLA-3 receptor) |
| 3655 | ITGA6 | integrin, alpha 6 |
| 3685 | ITGAV | integrin, alpha V (vitronectin receptor, alpha polypeptide, antigen CD51) |
| 3688 | ITGB1 | integrin, beta 1 (fibronectin receptor, beta polypeptide, antigen CD29 includes MDF2, MSK12) |
| 5747 | PTK2 | PTK2 protein tyrosine kinase 2 (EC:2.7.10.2) |
| 5293 | PIK3CD | phosphoinositide-3-kinase, catalytic, delta polypeptide (EC:2.7.1.153) |

| 5291 | PIK3CB | phosphoinositide-3-kinase, catalytic, beta polypeptide (EC:2.7.1.153) |
|---|---|---|
| 5294 | PIK3CG | phosphoinositide-3-kinase, catalytic, gamma polypeptide (EC:2.7.11.1 2.7.1.153) |
| 5290 | PIK3CA | phosphoinositide-3-kinase, catalytic, alpha polypeptide (EC:2.7.11.1 2.7.1.153) |
| 8503 | PIK3R3 | phosphoinositide-3-kinase, regulatory subunit 3 (gamma) |
| 23533 | PIK3R5 | phosphoinositide-3-kinase, regulatory subunit 5 |
| 5296 | PIK3R2 | phosphoinositide-3-kinase, regulatory subunit 2 (beta) |
| 5295 | PIK3R1 | phosphoinositide-3-kinase, regulatory subunit 1 (alpha) |
| 5728 | PTEN | phosphatase and tensin homolog (EC:3.1.3.67 3.1.3.16 3.1.3.48) |
| 4824 | NKX3-1 | NK3 homeobox 1 |
| 207 | AKT1 | v-akt murine thymoma viral oncogene homolog 1 (EC:2.7.11.1) |
| 208 | AKT2 | v-akt murine thymoma viral oncogene homolog 2 (EC:2.7.11.1) |
| 10000 | AKT3 | v-akt murine thymoma viral oncogene homolog 3 (protein kinase B, gamma) (EC:2.7.11.1) |
| 1147 | CHUK | conserved helix-loop-helix ubiquitous kinase (EC:2.7.11.10) |
| 3551 | IKBKB | inhibitor of kappa light polypeptide gene enhancer in B-cells, kinase beta (EC:2.7.11.10) |
| 8517 | IKBKG | inhibitor of kappa light polypeptide gene enhancer in B-cells, kinase gamma |
| 4792 | NFKBIA | nuclear factor of kappa light polypeptide gene enhancer in B-cells inhibitor, alpha |
| 4790 | NFKB1 | nuclear factor of kappa light polypeptide gene enhancer in B-cells 1 |
| 4791 | NFKB2 | nuclear factor of kappa light polypeptide gene enhancer in B-cells 2 (p49/p100) |
| 5970 | RELA | v-rel reticuloendotheliosis viral oncogene homolog A (avian) |
| 5743 | PTGS2 | prostaglandin-endoperoxide synthase 2 (prostaglandin G/H synthase and cyclooxygenase) (EC:1.14.99.1) |
| 4843 | NOS2 | nitric oxide synthase 2, inducible (EC:1.14.13.39) |
| 596 | BCL2 | B-cell CLL/lymphoma 2 |
| 330 | BIRC3 | baculoviral IAP repeat containing 3 |
| 331 | XIAP | X-linked inhibitor of apoptosis |
| 329 | BIRC2 | baculoviral IAP repeat containing 2 |
| 598 | BCL2L1 | BCL2-like 1 |
| 7185 | TRAF1 | TNF receptor-associated factor 1 |
| 7186 | TRAF2 | TNF receptor-associated factor 2 |
| 7187 | TRAF3 | TNF receptor-associated factor 3 |
| 9618 | TRAF4 | TNF receptor-associated factor 4 |
| 7188 | TRAF5 | TNF receptor-associated factor 5 |
| 7189 | TRAF6 | TNF receptor-associated factor 6, E3 ubiquitin protein ligase |
| 2475 | MTOR | mechanistic target of rapamycin (serine/threonine kinase) (EC:2.7.11.1) |

| 572 | BAD | BCL2-associated agonist of cell death |
|---|---|---|
| 2308 | FOXO1 | forkhead box O1 |
| 4193 | MDM2 | Mdm2, p53 E3 ubiquitin protein ligase homolog (mouse) (EC:6.3.2.19) |
| 7157 | TP53 | tumor protein p53 |
| 1027 | CDKN1B | cyclin-dependent kinase inhibitor 1B (p27, Kip1) |
| 1026 | CDKN1A | cyclin-dependent kinase inhibitor 1A (p21, Cip1) |
| 613 | BCR | breakpoint cluster region (EC:2.7.11.1) |
| 25 | ABL1 | c-abl oncogene 1, non-receptor tyrosine kinase (EC:2.7.10.2) |
| 1399 | CRKL | v-crk sarcoma virus CT10 oncogene homolog (avian)-like |
| 1398 | CRK | v-crk sarcoma virus CT10 oncogene homolog (avian) |
| 23624 | CBLC | Cbl proto-oncogene, E3 ubiquitin protein ligase C (EC:6.3.2.19) |
| 868 | CBLB | Cbl proto-oncogene, E3 ubiquitin protein ligase B (EC:6.3.2.19) |
| 867 | CBL | Cbl proto-oncogene, E3 ubiquitin protein ligase (EC:6.3.2.19) |
| 6776 | STAT5A | signal transducer and activator of transcription 5A |
| 6777 | STAT5B | signal transducer and activator of transcription 5B |
| 3716 | JAK1 | Janus kinase 1 (EC:2.7.10.2) |
| 6774 | STAT3 | signal transducer and activator of transcription 3 (acute-phase response factor) |
| 6772 | STAT1 | signal transducer and activator of transcription 1, 91kDa |
| 7423 | VEGFB | vascular endothelial growth factor B |
| 5228 | PGF | placental growth factor |
| 7422 | VEGFA | vascular endothelial growth factor A |
| 7424 | VEGFC | vascular endothelial growth factor C |
| 2277 | FIGF | c-fos induced growth factor (vascular endothelial growth factor D) |
| 7039 | TGFA | transforming growth factor, alpha |
| 1950 | EGF | epidermal growth factor |
| 1956 | EGFR | epidermal growth factor receptor (EC:2.7.10.1) |
| 2064 | ERBB2 | v-erb-b2 erythroblastic leukemia viral oncogene homolog 2, neuro/glioblastoma derived oncogene homolog (avian) (EC:2.7.10.1) |
| 5154 | PDGFA | platelet-derived growth factor alpha polypeptide |
| 5155 | PDGFB | platelet-derived growth factor beta polypeptide |
| 5156 | PDGFRA | platelet-derived growth factor receptor, alpha polypeptide (EC:2.7.10.1) |
| 5159 | PDGFRB | platelet-derived growth factor receptor, beta polypeptide (EC:2.7.10.1) |
| 3479 | IGF1 | insulin-like growth factor 1 (somatomedin C) |
| 3480 | IGF1R | insulin-like growth factor 1 receptor (EC:2.7.10.1) |
| 4254 | KITLG | KIT ligand |
| 3815 | KIT | v-kit Hardy-Zuckerman 4 feline sarcoma viral oncogene homolog (EC:2.7.10.1) |

| 2323 | FLT3LG | fms-related tyrosine kinase 3 ligand |
|---|---|---|
| 2322 | FLT3 | fms-related tyrosine kinase 3 (EC:2.7.10.1) |
| 3082 | HGF | hepatocyte growth factor (hepapoietin A |
| 4233 | MET | met proto-oncogene (hepatocyte growth factor receptor) (EC:2.7.10.1) |
| 2258 | FGF13 | fibroblast growth factor 13 |
| 2253 | FGF8 | fibroblast growth factor 8 (androgen-induced) |
| 8817 | FGF18 | fibroblast growth factor 18 |
| 2257 | FGF12 | fibroblast growth factor 12 |
| 2248 | FGF3 | fibroblast growth factor 3 |
| 2251 | FGF6 | fibroblast growth factor 6 |
| 2256 | FGF11 | fibroblast growth factor 11 |
| 8823 | FGF16 | fibroblast growth factor 16 |
| 2252 | FGF7 | fibroblast growth factor 7 |
| 2259 | FGF14 | fibroblast growth factor 14 |
| 8822 | FGF17 | fibroblast growth factor 17 |
| 9965 | FGF19 | fibroblast growth factor 19 |
| 2254 | FGF9 | fibroblast growth factor 9 (glia-activating factor) |
| 2250 | FGF5 | fibroblast growth factor 5 |
| 2249 | FGF4 | fibroblast growth factor 4 |
| 8074 | FGF23 | fibroblast growth factor 23 |
| 27006 | FGF22 | fibroblast growth factor 22 |
| 26281 | FGF20 | fibroblast growth factor 20 |
| 2255 | FGF10 | fibroblast growth factor 10 |
| 2247 | FGF2 | fibroblast growth factor 2 (basic) |
| 26291 | FGF21 | fibroblast growth factor 21 |
| 2246 | FGF1 | fibroblast growth factor 1 (acidic) |
| 2260 | FGFR1 | fibroblast growth factor receptor 1 (EC:2.7.10.1) |
| 2263 | FGFR2 | fibroblast growth factor receptor 2 (EC:2.7.10.1) |
| 2261 | FGFR3 | fibroblast growth factor receptor 3 (EC:2.7.10.1) |
| 2885 | GRB2 | growth factor receptor-bound protein 2 |
| 6654 | SOS1 | son of sevenless homolog 1 (Drosophila) |
| 6655 | SOS2 | son of sevenless homolog 2 (Drosophila) |
| 3265 | HRAS | v-Ha-ras Harvey rat sarcoma viral oncogene homolog |
| 3845 | KRAS | v-Ki-ras2 Kirsten rat sarcoma viral oncogene homolog |
| 4893 | NRAS | neuroblastoma RAS viral (v-ras) oncogene homolog |
| 369 | ARAF | v-raf murine sarcoma 3611 viral oncogene homolog (EC:2.7.11.1) |
| 673 | BRAF | v-raf murine sarcoma viral oncogene homolog B1 (EC:2.7.11.1) |
| 5894 | RAF1 | v-raf-1 murine leukemia viral oncogene homolog 1 (EC:2.7.11.1) |
| 5604 | MAP2K1 | mitogen-activated protein kinase kinase 1 (EC:2.7.12.2) |
| 5605 | MAP2K2 | mitogen-activated protein kinase kinase 2 (EC:2.7.12.2) |
| 5594 | MAPK1 | mitogen-activated protein kinase 1 (EC:2.7.11.24) |
| 5595 | MAPK3 | mitogen-activated protein kinase 3 (EC:2.7.11.24) |
| 3725 | JUN | jun proto-oncogene |
| 2353 | FOS | FBJ murine osteosarcoma viral oncogene homolog |
| 4312 | MMP1 | matrix metallopeptidase 1 (interstitial collagenase) (EC:3.4.24.7) |
| 4313 | MMP2 | matrix metallopeptidase 2 (gelatinase A, 72kDa gelatinase, 72kDa type IV collagenase) (EC:3.4.24.24) |
| 4318 | MMP9 | matrix metallopeptidase 9 (gelatinase B, 92kDa gelatinase, 92kDa type IV collagenase) (EC:3.4.24.35) |
| 3576 | IL8 | interleukin 8 |
| 1019 | CDK4 | cyclin-dependent kinase 4 (EC:2.7.11.22) |
| 5979 | RET | ret proto-oncogene (EC:2.7.10.1) |
| 8030 | CCDC6 | coiled-coil domain containing 6 |
| 8031 | NCOA4 | nuclear receptor coactivator 4 |
| 4914 | NTRK1 | neurotrophic tyrosine kinase, receptor, type 1 (EC:2.7.10.1) |
| 7170 | TPM3 | tropomyosin 3 |
| 7175 | TPR | translocated promoter region (to activated MET oncogene) |
| 10342 | TFG | TRK-fused gene |
| 11186 | RASSF1 | Ras association (RalGDS/AF-6) domain family member 1 |
| 83593 | RASSF5 | Ras association (RalGDS/AF-6) domain family member 5 |
| 6789 | STK4 | serine/threonine kinase 4 (EC:2.7.11.1) |
| 1613 | DAPK3 | death-associated protein kinase 3 (EC:2.7.11.1) |
| 23604 | DAPK2 | death-associated protein kinase 2 (EC:2.7.11.1) |
| 1612 | DAPK1 | death-associated protein kinase 1 (EC:2.7.11.1) |
| 5335 | PLCG1 | phospholipase C, gamma 1 (EC:3.1.4.11) |
| 5336 | PLCG2 | phospholipase C, gamma 2 (phosphatidylinositol-specific) (EC:3.1.4.11) |
| 5579 | PRKCB | protein kinase C, beta (EC:2.7.11.13) |
| 5578 | PRKCA | protein kinase C, alpha (EC:2.7.11.13) |
| 5582 | PRKCG | protein kinase C, gamma (EC:2.7.11.13) |
| 5900 | RALGDS | ral guanine nucleotide dissociation stimulator |
| 5898 | RALA | v-ral simian leukemia viral oncogene homolog A (ras related) |
| 5899 | RALB | v-ral simian leukemia viral oncogene homolog B (ras related |
| 10928 | RALBP1 | ralA binding protein 1 |
| 998 | CDC42 | cell division cycle 42 (GTP binding protein, 25kDa) |
| 5879 | RAC1 | ras-related C3 botulinum toxin substrate 1 (rho family, small GTP binding protein Rac1) |
| 5880 | RAC2 | ras-related C3 botulinum toxin substrate 2 (rho family, small GTP binding protein Rac2) |
| 5881 | RAC3 | ras-related C3 botulinum toxin substrate 3 (rho family, small GTP binding protein Rac3) |
| 387 | RHOA | ras homolog family member A |
| 5602 | MAPK10 | mitogen-activated protein kinase 10 (EC:2.7.11.24) |
| 5601 | MAPK9 | mitogen-activated protein kinase 9 (EC:2.7.11.24) |
| 5599 | MAPK8 | mitogen-activated protein kinase 8 (EC:2.7.11.24) |
| 7849 | PAX8 | paired box 8 |
| 5468 | PPARG | peroxisome proliferator-activated receptor gamma |
| 6256 | RXRA | retinoid X receptor, alpha |
| 6257 | RXRB | retinoid X receptor, beta |
| 6258 | RXRG | retinoid X receptor, gamma |
| 5915 | RARB | retinoic acid receptor, beta |
| 5467 | PPARD | peroxisome proliferator-activated receptor delta |
| 3728 | JUP | junction plakoglobin |
| 7704 | ZBTB16 | zinc finger and BTB domain containing 16 |
| 5371 | PML | promyelocytic leukemia |
| 5914 | RARA | retinoic acid receptor, alpha |
| 861 | RUNX1 | runt-related transcription factor 1 |
| 862 | RUNX1T1 | runt-related transcription factor 1 |
| 6688 | SPI1 | spleen focus forming virus (SFFV) proviral integration oncogene spi1 |
| 1050 | CEBPA | CCAAT/enhancer binding protein (C/EBP), alpha |
| 1438 | CSF2RA | colony stimulating factor 2 receptor, alpha, low-affinity (granulocyte-macrophage) |
| 1441 | CSF3R | colony stimulating factor 3 receptor (granulocyte) |
| 1436 | CSF1R | colony stimulating factor 1 receptor (EC:2.7.10.1) |
| 3569 | IL6 | interleukin 6 (interferon, beta 2) |
| 1029 | CDKN2A | cyclin-dependent kinase inhibitor 2A (melanoma, p16, inhibits CDK4) |
| 1871 | E2F3 | E2F transcription factor 3 |
| 1869 | E2F1 | E2F transcription factor 1 |
| 1870 | E2F2 | E2F transcription factor 2 |
| 4149 | MAX | MYC associated factor X |
| 8554 | PIAS1 | protein inhibitor of activated STAT, 1 |
| 51588 | PIAS4 | protein inhibitor of activated STAT, 4 |
| 10401 | PIAS3 | protein inhibitor of activated STAT, 3 |
| 9063 | PIAS2 | protein inhibitor of activated STAT, 2 |
| 1030 | CDKN2B | cyclin-dependent kinase inhibitor 2B (p15, inhibits CDK4) |
| 1021 | CDK6 | cyclin-dependent kinase 6 (EC:2.7.11.22) |
| 1163 | CKS1B | CDC28 protein kinase regulatory subunit 1B |
| 6502 | SKP2 | S-phase kinase-associated protein 2, E3 ubiquitin protein ligase |
| 1017 | CDK2 | cyclin-dependent kinase 2 (EC:2.7.11.22) |
| 9134 | CCNE2 | cyclin E2 |
| 898 | CCNE1 | cyclin E1 |
| 5925 | RB1 | retinoblastoma 1 |
| 4286 | MITF | microphthalmia-associated transcription factor |
| 7040 | TGFB1 | transforming growth factor, beta 1 |
| 7042 | TGFB2 | transforming growth factor, beta 2 |
| 7043 | TGFB3 | transforming growth factor, beta 3 |
| 7046 | TGFBR1 | transforming growth factor, beta receptor 1 (EC:2.7.11.30) |
| 7048 | TGFBR2 | transforming growth factor, beta receptor II (70/80kDa) (EC:2.7.11.30) |
| 4087 | SMAD2 | SMAD family member 2 |
| 4088 | SMAD3 | SMAD family member 3 |
| 4089 | SMAD4 | SMAD family member 4 |

| 2122 | MECOM | MDS1 and EVI1 complex locus |
|---|---|---|
| 1488 | CTBP2 | C-terminal binding protein 2 |
| 1487 | CTBP1 | C-terminal binding protein 1 |
| 3066 | HDAC2 | histone deacetylase 2 (EC:3.5.1.98) |
| 3065 | HDAC1 | histone deacetylase 1 (EC:3.5.1.98) |
| 4292 | MLH1 | mutL homolog 1, colon cancer, nonpolyposis type 2 (E. coli) |
| 4436 | MSH2 | mutS homolog 2, colon cancer, nonpolyposis type 1 (E. coli) |
| 4437 | MSH3 | mutS homolog 3 (E. coli) |
| 2956 | MSH6 | mutS homolog 6 (E. coli) |
| 581 | BAX | BCL2-associated X protein |
| 675 | BRCA2 | breast cancer 2, early onset |
| 5888 | RAD51 | RAD51 homolog (S. cerevisiae) |
| 356 | FASLG | Fas ligand (TNF superfamily, member 6) |
| 355 | FAS | Fas (TNF receptor superfamily, member 6) |
| 8772 | FADD | Fas (TNFRSF6)-associated via death domain |
| 841 | CASP8 | caspase 8, apoptosis-related cysteine peptidase (EC:3.4.22.61) |
| 637 | BID | BH3 interacting domain death agonist |
| 54205 | CYCS | cytochrome c, somatic |
| 7428 | VHL | von Hippel-Lindau tumor suppressor, E3 ubiquitin protein ligase |
| 6921 | TCEB1 | transcription elongation factor B (SIII), polypeptide 1 (15kDa, elongin C) |
| 6923 | TCEB2 | transcription elongation factor B (SIII), polypeptide 2 (18kDa, elongin B) |
| 9978 | RBX1 | ring-box 1, E3 ubiquitin protein ligase |
| 8453 | CUL2 | cullin 2 |
| 112399 | EGLN3 | egl nine homolog 3 (C. elegans) (EC:1.14.11.29) |
| 112398 | EGLN2 | egl nine homolog 2 (C. elegans) (EC:1.14.11.29) |
| 54583 | EGLN1 | egl nine homolog 1 (C. elegans) (EC:1.14.11.29) |
| 2271 | FH | fumarate hydratase (EC:4.2.1.2) |
| 3091 | HIF1A | hypoxia inducible factor 1, alpha subunit (basic helix-loop-helix transcription factor) |
| 2034 | EPAS1 | endothelial PAS domain protein 1 |
| 405 | ARNT | aryl hydrocarbon receptor nuclear translocator |
| 9915 | ARNT2 | aryl-hydrocarbon receptor nuclear translocator 2 |
| 1387 | CREBBP | CREB binding protein (EC:2.3.1.48) |
| 2033 | EP300 | E1A binding protein p300 (EC:2.3.1.48) |
| 6513 | SLC2A1 | solute carrier family 2 (facilitated glucose transporter), member 1 |
| 6469 | SHH | sonic hedgehog |
| 5727 | PTCH1 | patched 1 |
| 6608 | SMO | smoothened, frizzled family receptor |
| 27148 | STK36 | serine/threonine kinase 36 (EC:2.7.11.1) |
| 51684 | SUFU | suppressor of fused homolog (Drosophila) |
| 2737 | GLI3 | GLI family zinc finger 3 |

| 2736 | GLI2 | GLI family zinc finger 2 |
| 2735 | GLI1 | GLI family zinc finger 1 |
| 650 | BMP2 | bone morphogenetic protein 2 |
| 652 | BMP4 | bone morphogenetic protein 4 |
| 64399 | HHIP | hedgehog interacting protein |
| 8643 | PTCH2 | patched 2 |
| 367 | AR | androgen receptor |
| 3320 | HSP90AA1 | heat shock protein 90kDa alpha (cytosolic), class A member 1 |
| 3326 | HSP90AB1 | heat shock protein 90kDa alpha (cytosolic), class B member 1 |
| 7184 | HSP90B1 | heat shock protein 90kDa beta (Grp94), member 1 |
| 354 | KLK3 | kallikrein-related peptidase 3 (EC:3.4.21.77) |
| 112401 | BIRC8 | baculoviral IAP repeat containing 8 |
| 2113 | ETS1 | v-ets erythroblastosis virus E26 oncogene homolog 1 (avian) |
| 2950 | GSTP1 | glutathione S-transferase pi 1 (EC:2.5.1.18) |
| 5337 | PLD1 | phospholipase D1, phosphatidylcholine-specific (EC:3.1.4.4) |
| 79444 | BIRC7 | baculoviral IAP repeat containing 7 |
| 8900 | CCNA1 | cyclin A1 |

**Table S1.3. SBCGs (391) compiled for metastasis of primary breast and prostate cancer to bone.**

| SBCGs compiled for metastasis of primary breast cancer to bone | |
|---|---|
| **Gene Symbol** | **Reference** |
| LRRC15, MMP13, MMP2, BCAM, COL10A1, AEBP1, IGFBP4, ESR1, AEBP1, ASPN, ATP6V0C, BCAM, COMP, CRIP1, DPT, IGFBP4, LRRC15, LRP1, MAB21L2, MMP13, MMP2, PDLIM7, PTOV1, PCDHGC3, RFNG, STK32B, TBC1D10B, TCIRG1, ZNF444 | [1] |
| COX2, PGE2, TNFSF11, TNFRSF11A, FGF7, PTX3, NID2, RAP1GDS1, NPNT, COL6A3, IFIT3, IGRC, FMOD, IGIT1B, EGR1, PTGS2, CXCL10, VCAM1, FGF7, PTN, CD1D, LAMB1, CCL5, SFRP2, CXCL1 | [2] |
| RASSF1A, MGMT, RAR-β2, APC | [3] |
| IBSP, SPP1, FGF13, COX7B2, SPARC, PRG1, GJA1, SOCS2, EDN1, IL11, CYR61, IGFBP4, IL8, PLAU, FST, MADH3, PTGS2, EXTL2, CTGF, CBFB, CDH11, S100A4, TIMP3, NOG, TGFA, THBS1, IGFBP3, PLAT, TIMP2, FN1, CCNB1, LRPAP1, MMP1, COL5A1, COL8A1, CALB2, CYR61, CTGF, WISP, BMP2, IL11, FST, CXCR4, ADAMTS1, MMP1, CX43, DUSP1, NOG, COX2, CBFB, GJA1, THBS1 | [4] |
| SERPINB5 | [5] |
| ABCB1, ESR1, RUNDC3B, RAP2A | [6] |
| CSF1R | [7] |
| CAV1 | [8] |
| TGFB1, ATF3, MMP13, CCNA1 | [9] |
| COL1A1, VIM, S100A4, ACTA2, PDGFRB, NG2, CXCL1, CXCL5, PTGS2 | [10] |
| HISTIH2AC, POMZP3, PON2, TMC5, SCNN1A, C10orf116, NAP1L3, GPRD5C, CTGF, FHL1, DUSP1, DLC1, SOCS2, IL11, ADAMTS1, CYP1B1, SAA2, SAA1, PPL, IL11, SPANXB1, FST, SERPINA1, HLA-DRA, HLA-DPA1, ABCC3, CSGALNACT1, FKBP11, CXCR4, PRG1, SES7-1, SOX4, KHDRBS3, FGF5, C14orf1139, PTX7, MCAM, TGFBI, S100A2, MMP1, FGF5, RGC32 | [11] |
| TFF1, TFF3, AGR2, NAT1, CRIP1, TNRC9, SCUBE2, TNRC9, CYP2B6, RND1, DLALT1, KIF5C, PLA2GHB, UNG2, HMGCS2, SLC1A1, CEACAM6, TSPAN-1, REPS2, HPX, PDE4DIP, TOM1L1, SCGB2A2, ANXA9, BCAS1, TIMP4, CYP2B6, C9orf116, MSMB, FGFR3, FGFBP1, BAG1, FOXO3A, KRT16, MALL, KCNG1, IGLC2, KLK8, KRT6B, SNAI1, TMSNB, NRTN, EPHB3, ENO1, SOD2, IGHG1, KLK5, SOS, RARRES1, TTYH1, CD24, MGC27165, SERPINB5, UCHL1, ROPN1, LOC56901, ITGA6, TUBB, UBE4B, IGHM, COL2A1, IMPA2, DRE1, KLK7, TAZ, USP34, C6orf4, ALAD, ARHGDIA, ARS2, CLCA2, ELMO2, LRRC31, HTATSF1, HTR2B, MLPH, SLC2A8, TFPI2, BRCA1, BRCA2 | [12] |
| DIP2B, MCM6, PLEK2 , FOXI1 , SYT13 , PRDM6 , LONP2, ZNF185, ENTPD4, PPP2R2C, LANCL2, SFTPD, TSGA10, B3GALT2, SEZ6L, KCND2, VSTM2, FANCL, FDPS , VRK2 | [13] |
| HIF1A, VEGF, DUSP1, CXCR4 | [14] |
| IL1B, COL1A1, COL1A2, CTGF, IL10, IL8, SPP1, VCAM1, PGE2, EDN1, BMP6, IL6, PDGFRA, PTHLH, TGFB, LAMC2, IGFBP3, PLAU, TNF, TNFSF11, TNFRSF11A, TNFRSF11B, CXCR4, | [15] |
| CXCL12 | |
| CXCR4, CXCL12, HER2 | [16] |
| PASAT1, PHGDH, PSPH, LTBP1, ATF3, CCNA1, MMP13 | [17] |
| ADAM9, ADAMTS1, AREG, EGF, EGFR, ERBB2, HBEGF, MMP1, PTHLH, TNFRSF11A, TNFRSF11B | [18] |
| CTNNB1, DKK1, PLAU, PLAUR, SERPINB2, SERPINE1, WNT3A | [19] |
| BMP4, FGF1, FGF17, FGF4, FGF6, FGF8, IGF1, IL11, IL1B, IL6, IL8 | [20] |
| CTTN, CXCL12, CXCR4, IL11, IL6, | [21] |
| ADAMTS1, EGF, EGFR, MMP1, TNFRSF11A, TNFRSF11B | [22] |
| HIF1A | [23] |
| AKT1, AKT3, BMP2, CLEC11A, CXCL12, CXCR4, JAG1, NOV, PDGFA, PGF, PRG2, SPP1, SRC, TGFB1, TGFB3, TNFSF10, VEGFC | [24] |
| CTGF, NOV, WISP2, WISP3 | [25] |
| CSF1, IL8, PTHLH, TNFRSF11A, TNFRSF11B | [26] |
| **SBCGs compiled for metastasis of primary prostate cancer to bone** | |
| APX1, ANTXR1, CDC23, CDC37, CETN3, CENPE, CCNC, CCNG1, CDK4, CDKN2C, CDKN3, CDKL1, DDB1, E2F1, FEN1, GTPBP4, KAT7, MCM7, MAD2L1, MSH6, NBN, PCNA, EIF2AK2, RAN, RPA2, RPA3, RRM2, SSBP1, SKP1, ADD1, DST, CDH1, CTNNAL1, CYB5R3, DSC1, DSG2, ADAM9, FER, TNC, ITGB1, MFGE8, MCAM, ACVRL1, BGN, BST1, IBSP, CDH11, COL1A2, COL6A1, COL6A2, COL7A1, COL11A1, COL16A1, CSF1, DCN, FSTL1, INHA, MGP, SHH, VCAN, COL17A1, ITGB4, PLEC, DSC2, DSP, PKP1, PKP2, PKP3, PKP4, ASRGL1, BMP2, VDR, RUNX2, BGLAP, SPARC, SPP1, TNFRSF11B, TNFSF11, DRG1, PTHRP, EDN1, TNFRSF11A, TP53, HPN, TFF3, BGLAP, ERBB3, PSMD9, CDKN1A, MYCN, AR, HIF1A, MAGEA1, VEGF, KLK3, GDF15, BMP6, ERBB2, LTBP1, MMP9, MMP1, MMP3, PLAU, HRAS, NRAS, KRT1, PTRF, MDM2, RB1 | [27] |
| HPN | [28] |
| TFF3 | [29] |
| BGLAP | [30] |
| ERBB3 | [31] |
| PSMD9, CDKN1A | [27] |
| ERBB2, EGFR | [32] |
| PALB, TGFB1, EGFR, BMP | [33] |
| BIN1, MYC, ABL1 | [34] |
| MAGEA1 | [35] |
| MMP1, MMP2, MMP13, DRG1, PTEN, NME1, CD82, KISS1, BRMS1, MAP2K4, TP53 | [36] |
| IGFBP3, PTH1R, TNFSF11, MMP, MDM2, RB1, TNFRSF11A | [37] |
| RB1 | [38] |
| CDH11 | [39] |
| LYN, SRC | [40] |
| ADAMTS1, AREG, EGF, EGFR, HBEGF, PTHLH, TGFA, TNF, TNFRSF11A, TNFRSF11B, VEGFA | [18] |
| VEGFA, NR1I2 | [41] |
| AKT1, PTHLH, TNFRSF11A, TNFRSF11B | [42] |
| BMP4, CSF1, DKK1, IL8, NOG, PTHLH, SFRP1, SFRP2, TGFB1, TNFRSF11A, TNFRSF11B | [43] |
| RAP1GAP, TERF2IP, VTNR | [44] |
| CXCL12, CXCR4, MAPK1, MAPK3, | [45] |
| FGF1, FGF17, FGF4, FGF6, FGF8, FGFR3, | [46] |

**References (for Supplementary Table S1.3)**

**Table S1.4. Significantly enriched GO terms, characteristic to metastasis to bone, identified from enrichment analysis of SBCGs.**

| GO ID | GO terms | Reference |
|---|---|---|
| **Bone related processes** | | |
| GO:0001958 | endochondral ossification | [1] |
| GO:0036075 | replacement ossification | [1] |
| GO:0030282 | bone mineralization | [1] |
| GO:0030199 | collagen fibril organization | [2] |
| GO:0032963 | collagen metabolic process | [2] |
| GO:0033280 | response to vitamin D | [3] |
| GO:0046850 | regulation of bone remodeling | [4] |
| GO:0001649 | osteoblast differentiation | [1,4] |
| GO:0001503 | ossification | [1] |
| GO:0051216 | cartilage development | [1] |
| **Metastasis related processes** | | |
| GO:0002690 | positive regulation of leukocyte chemotaxis | [5] |
| GO:0002688 | regulation of leukocyte chemotaxis | [5] |
| GO:0002687 | positive regulation of leukocyte migration | [5] |
| GO:0050920 | regulation of chemotaxis | [5] |
| GO:0007162 | negative regulation of cell adhesion | [6] |
| GO:0002685 | regulation of leukocyte migration | [7] |
| GO:0050900 | leukocyte migration | [7] |
| GO:0016337 | cell-cell adhesion | [6,8] |
| GO:0001525 | angiogenesis | [9,10] |
| GO:0050679 | positive regulation of epithelial cell proliferation | [3] |
| GO:0045785 | positive regulation of cell adhesion | [6] |
| GO:0043236 | laminin binding | [11] |
| GO:0001968 | fibronectin binding | [12,13] |
| GO:0005104 | fibroblast growth factor receptor binding | [14] |
| GO:0048407 | platelet-derived growth factor binding | [15,16] |
| GO:0005518 | collagen binding | [2,17] |
| GO:0005178 | integrin binding | [12,18] |
| GO:0005539 | glycosaminoglycan binding | [19] |
| GO:0005125 | cytokine activity | [20,21] |
| GO:0030246 | carbohydrate binding | [22] |
| GO:0035413 | positive regulation of catenin import into nucleus | [23] |

**References (for Supplementary Table S1.4)**

**Table S1.5. Relevance of SBC Targets.**

| SBC Specific Targets | Reference |
|---|---|
| TNXB | [1] |
| SPP1 | [2] |
| CTGF | [3–5] |
| BMP1 | [6] |
| BMPR1A | [6] |
| CD44 | [7] |
| VWF | [8] |

**References (for Supplementary Table S1.5)**